\documentclass{emulateapj}
\usepackage{graphicx,amsmath,amsfonts,amssymb,subfigure,hyperref,enumerate}

\newcommand{\kms}{\,~km~s$^{-1}$\ }

\newcommand{\Mpc}{\>{\rm Mpc}}

\newcommand{\ALFALFA}{ALFALFA40}

\def\MedianGasFractionIso{0.81 \pm  0.13} 
\def\MedianGasFractionSats{0.82 \pm  0.16} 
\def\NumberLowMassSatsWithGas{429} 
\def\NumberLowMassIsoWithGas{368} 
\def\NumberLowMassAllWithGas{797} 
\def\IsolatedFractionWithGas{0.46} 
\def\LowMassThreshWithGas{9.0}
\def\OurMinHIMass{6.8}
\def\ALFALFAMinHIMass{7.2}
\def\NAOverlap{15} 
\def\mhimstarbreak{8.6}
\def\mhimstarlowslope{1.052}
\def\mhimstarlowslopeerr{0.058}
\def\mhimstarlowconst{0.236}
\def\mhimstarlowconsterr{0.476}
\def\mhimstarlowscatter{0.285}
\def\mhimstarlowscattererr{0.019}
\def\mhimstarhighslope{0.461}
\def\mhimstarhighslopeerr{0.011}
\def\mhimstarhighconst{5.329}
\def\mhimstarhighconsterr{0.112}
\def\mhimstarhighscatter{0.221}
\def\mhimstarhighscattererr{0.006}
\def\NHIALFALFA{4098} 
\def\NALFALFARCUT{2.0} 
\def\NHIALFALFACUT{169} 
\def\ALFALFATWOSIGPERC{96} 
\def\NHIALFALFAHIQ{4032} 
\def\NHIOurs{255} 
\def\NNewIso{148} 
\def\NTOTALREDUCED{371}
\def\NTOTALHIGHQUALITY{317}
\def\NTOTALNONDETECT{54}
\def\NTOTALBADBASELINES{23}

\def\NTOTALISOLATED{144}
\def\NTOTALISOND{7}
\def\isovmslope{3.238}
\def\isovmslopeerr{0.046}
\def\isovmconst{3.210}
\def\isovmconsterr{0.097}
\def\isovmscatter{0.250}
\def\isovmscattererr{0.008}
\def\isovmpearsonrank{0.916}
\def\isomvslope{0.277}
\def\isomvslopeerr{0.004}
\def\isomvconst{-0.672}
\def\isomvconsterr{0.041}
\def\isomvscatter{0.075}
\def\isomvscattererr{0.002}
\def\isomvpearsonrank{0.916}
\def\isovrslope{1.245}
\def\isovrslopeerr{0.033}
\def\isovrconst{-2.005}
\def\isovrconsterr{0.069}
\def\isovrscatter{0.209}
\def\isovrscattererr{0.005}
\def\isovrpearsonrank{0.777}

\def\isoMrslope{0.406}
\def\isoMrslopeerr{0.007}
\def\isoMrconst{-3.456}
\def\isoMrconsterr{0.074}
\def\isoMrscatter{0.159}
\def\isoMrscattererr{0.004}
\def\isoMrpearsonrank{0.880}
\def\isofbaryonvpower{0.435}
\def\nonvmslope{2.949}
\def\nonvmslopeerr{0.053}
\def\nonvmconst{3.834}
\def\nonvmconsterr{0.112}
\def\nonvmscatter{0.261}
\def\nonvmscattererr{0.007}
\def\nonvmpearsonrank{0.831}
\def\nonmvslope{0.266}
\def\nonmvslopeerr{0.005}
\def\nonmvconst{-0.565}
\def\nonmvconsterr{0.050}
\def\nonmvscatter{0.080}
\def\nonmvscattererr{0.002}
\def\nonmvpearsonrank{0.831}
\def\nonvrslope{1.020}
\def\nonvrslopeerr{0.039}
\def\nonvrconst{-1.522}
\def\nonvrconsterr{0.083}
\def\nonvrscatter{0.225}
\def\nonvrscattererr{0.005}
\def\nonvrpearsonrank{0.577}

\def\nonMrslope{0.390}
\def\nonMrslopeerr{0.009}
\def\nonMrconst{-3.293}
\def\nonMrconsterr{0.094}
\def\nonMrscatter{0.179}
\def\nonMrscattererr{0.004}
\def\nonMrpearsonrank{0.764}

\def\allvmslope{3.113}
\def\allvmslopeerr{0.035}
\def\allvmconst{3.482}
\def\allvmconsterr{0.074}
\def\allvmscatter{0.258}
\def\allvmscattererr{0.006}
\def\allvmpearsonrank{0.878}
\def\allmvslope{0.272}
\def\allmvslopeerr{0.003}
\def\allmvconst{-0.624}
\def\allmvconsterr{0.031}
\def\allmvscatter{0.078}
\def\allmvscattererr{0.002}
\def\allmvpearsonrank{0.878}
\def\allvrslope{1.147}
\def\allvrslopeerr{0.026}
\def\allvrconst{-1.793}
\def\allvrconsterr{0.054}
\def\allvrscatter{0.219}
\def\allvrscattererr{0.004}
\def\allvrpearsonrank{0.683}

\def\allMrslope{0.400}
\def\allMrslopeerr{0.006}
\def\allMrconst{-3.386}
\def\allMrconsterr{0.059}
\def\allMrscatter{0.171}
\def\allMrscattererr{0.003}
\def\allMrpearsonrank{0.826}

\def\FBARYONDISKFRACTION{0.15 \pm 0.17} 
\def\FBARYONHALOFRACTIONVDB{0.04 \pm 0.06} 
\def\FBARYONHALOFRACTIONB{0.04 \pm 0.02} 
\def\NALLINNSA{106374}
\def\NISOINNSA{36592}
\def\NNONISOINNSA{69782}
\def\FRACISONSA{0.34}
\def\ALFALFAMedianColor{0.5}
\def\OurMedianColor{0.3}
\begin{document}
\title{A study in blue: the baryon content of isolated low mass galaxies}
\author{Jeremy D. Bradford, Marla C. Geha}
\affil{Astronomy Department, Yale University, New Haven, CT~06520,
 jeremy.bradford@yale.edu, marla.geha@yale.edu}

\author{Michael R. Blanton}
\affil{Center for Cosmology and Particle Physics, New York
 University, New York, NY~10003, blanton@nyu.edu}

\begin{abstract}
\renewcommand{\thefootnote}{\fnsymbol{footnote}}

We study the baryon content of low mass galaxies selected from the Sloan Digital Sky Survey (SDSS DR8), focusing on galaxies in isolated environments where the complicating physics of galaxy-galaxy interactions are minimized. We measure neutral hydrogen (HI) gas masses and line-widths for \NNewIso~isolated galaxies with stellar mass between $10^7$ and $10^{9.5} M_{\odot}$. We compare isolated low mass galaxies to more massive galaxies and galaxies in denser environments by remeasuring HI emission lines from the Arecibo Legacy Fast ALFA (ALFALFA) survey 40\% data release. All isolated low mass galaxies either have large atomic gas fractions or large atomic gas fractions cannot be ruled out via their upper limits. We measure a median atomic gas fraction of $f_{\rm gas} = \MedianGasFractionIso$ for our isolated low mass sample with no systems below 0.30. At all stellar masses, the correlations between galaxy radius, baryonic mass and velocity width are not significantly affected by environment. Finally, we estimate a median baryon to total dynamical mass fraction of $f_{\rm baryon, disk} = \FBARYONDISKFRACTION$. We also estimate two different median baryon to halo mass fractions using the results of semi-analytic models $(f_{\rm baryon, halo} = \FBARYONHALOFRACTIONVDB)$ and abundance matching $(f_{\rm baryon, halo} = \FBARYONHALOFRACTIONB)$. Baryon fractions estimated directly using HI observations appear independent of environment and maximum circular velocity, while baryon fractions estimated using abundance matching show a significant depletion of baryons at low maximum circular velocities.

\end{abstract}


\section{Introduction}
\label{sec_intro}

Environment plays a major in role in the evolution and the observed properties of galaxies \citep[e.g.,][]{Dressler:1980ie, Kauffmann:2004cw, Blanton:2005eb, Peng:2010gn, Geha:2012eu, Tal:2014uf, Wetzel:2014ju}. Environmental processes such as tidal forces and ram pressure stripping shape, quench and destroy galaxies \citep[e.g.,][]{Toomre:1972ji, Gunn:1972gx, Moore:1996il, Penarrubia:2008eq}. These processes will be especially pronounced in low mass galaxies with stellar masses below $10^9 M_{\odot}$, where more massive nearby galaxies dramatically affect the local gravitational potential \citep[e.g.,][]{Kravtsov:2004he, Zolotov:2012hi, Kenney:2013bf}. 

The effects of environment complicate both observational interpretations and comparisons to theoretical predictions of galaxy formation. Galaxies in isolation are {\it relatively} undisturbed by environmental processes. Thus, isolated galaxies can potentially present a clearer picture of galaxy evolution \citep{Karachentseva:1973we, Barton:2007be, Cortese:2011hz, Karachentsev:2011ju, Tollerud:2011jn, Toribio:2011eb}. Recent work has found that the properties of isolated low mass galaxies, loosely defined as ``field galaxies", appear to conflict with small-scale predictions of $\Lambda$CDM cosmology \citep{Kirby:2014gk, Klypin:2014ue, Papastergis:2015bc}. It is critical that isolated low mass galaxies are clearly defined and well studied. 

Isolated galaxies retain their HI gas out to larger radii \citep[e.g.,][]{Begum:2008cx}, contain a large fraction of baryons in atomic gas \citep[e.g.,][hereafter G06]{Grcevich:2009fs,Geha:2006jx} and are predicted to exhibit larger maximum circular velocities than galaxies that have been tidally stripped \citep[e.g.,][]{Brooks:2014jv}. Therefore in underdense regions, where galaxies are less affected by environmental processes, it might be easier to predict their properties when compared to galaxies in dense environments \citep{McGaugh:2010ky, Cortese:2011hz, Huang:2012gk, ArgudoFernandez:2014cq}.

Here we present the largest homogeneously measured data set of baryons in isolated low mass galaxies with $10^{7} M_{\odot} < M_{*} < 10^{9.5} M_{\odot}$. Our new data probe isolated galaxies with average circular velocities of 45\kms and average baryon masses of $2 \times 10^{8}~M_{\odot}$, with circular velocities as low as $\sim 20$\kms and baryon masses as low as $\sim 10^{7.4} M_{\odot}$. We complement the existing ALFALFA 40\% data release of \citet{Haynes:2011en}, hereafter \ALFALFA, with new 21 cm observations, incorporating our isolation criterion into the selection of \NNewIso~low mass targets.

This paper is organized as follows: In \S\,\ref{sec_data}, we introduce the data, define our isolation criteria and describe our automated HI parameter estimates. In \S\,\ref{sec_results} we present our results, examining the relationship between galaxy stellar mass, atomic gas mass and atomic gas fraction (\S\,\ref{subsec_m_star_f_gas} and \ref{subsec_m_star_m_gas}), the relationship between galaxy size, baryonic mass and dynamical mass (\S\,\ref{subsec_vms_scaling}), and finally baryon fractions in \S\,\ref{subsec_baryon_fractions}. In \S\,\ref{sec_discussion} we discuss our results in context with the literature, particularly with respect to the effects of environment on galaxy properties. Finally, we summarize our findings in \S\,\ref{sec_summary}.

Throughout this work we assume the following cosmological parameters: $\Omega_0 = 0.3$, $\Omega_{\Lambda} = 0.7$, $H_0 = 70$\kms (i.e., h = 0.7), and a cosmic baryon fraction of $f_b = 0.17$ \citep{Komatsu:2011in}. We present a small sample of isolated low mass galaxies for reference in Table \ref{tbl_hienv}. Since our HI observations are ongoing, our complete data set and raw HI spectra are available upon request. This paper is the first in a series using isolated low mass galaxies as a control sample for testing galaxy evolution at small scales.

\section{Data} 
\label{sec_data}

\subsection{Galaxy Catalog}
\label{subsec_nsa}

To construct a sample of isolated low mass galaxies, we begin with the NASA-Sloan Atlas\footnote{\url{http://www.nsatlas.org}} (NSA) catalog \citep{Blanton:2011dv}, which is itself a re-reduction of the SDSS Data Release 8 \citep[DR8, ][]{Aihara:2011kj}. This sample was first described in \citet[][hereafter G12]{Geha:2012eu}, and we provide a brief description here. 

The NSA re-reduction is optimized for nearby extended objects with $z < 0.055$ \citep{Blanton:2011dv}. The NSA photometry includes image mosaics from both SDSS \textit{ugriz} and GALEX FUV, NUV bands \citep{Martin:2005ko}. S\'ersic model structural parameters are fit using the SDSS r-band image. The SDSS spectra are also reanalyzed and recalibrated \citep{Yan:2011in, Yan:2012jn}. The spectra equivalent widths are re-measured and included in the NSA catalog. 

In order to ensure accurate distance estimates and to stay within the spectroscopic completeness limits of the SDSS, we discard any galaxy in the NSA catalog where it is not detected in the SDSS $g$, $r$ or $i$ passbands and only include galaxies with spectroscopy from the main SDSS legacy survey with $z > 0.002$ and $M_{r} < 17.72$. We also discard galaxies where the S\'ersic half-light radius is larger than the Petrosian 90\% light radius, which indicates that surface brightness model fitting has failed. Our final catalog is comprised of \NALLINNSA~galaxies that span distances of $10 < D < 250~\Mpc$ and stellar masses of $10^{6.1} M_{\odot} < M_* < 10^{11.9} M_{\odot}$.

Galaxy stellar masses ($M_{*}$) are calculated in the NSA catalog using the \texttt{kcorrect} software of \citet{Blanton:2007kl} using the SDSS and GALEX photometric bands and assuming a \citet{Chabrier:2003ki} IMF. The error in stellar mass is estimated from the scatter between models examined by \citet{Blanton:2007kl} who found roughly uniform differences of 0.2 dex between models (see their Fig.~17). We calculate luminosity distances using SDSS redshifts corrected for peculiar velocities using the model of \citet{Willick:1997gg}, who compiled a velocity-distance catalog of spiral galaxies using Tully-Fisher calibrated distances and redshift velocities. Distance uncertainties are folded into all distant dependent calculations.
\subsection{Galaxy Environments, Isolation Criteria and Sample Definition}
\label{subsec_env}

Many methods have been devised for quantifying the local and large-scale environments around galaxies. To measure environment density and degree of isolation, it is common to define an environment variable, such as nth nearest neighbor surface density \citep[e.g.,][]{Brough:2013df} or tidal force index \citep[e.g.,][]{Karachentsev:2011ju}. It is also common to define galaxies ``in the field" in order to study galaxies in isolation, where such a galaxy has various definitions in the literature. Other studies measure the density of galaxies within a fixed-aperture on the sky or measure 2D projected distances between galaxies with cuts applied in redshift space \citep[e.g.,][G12]{Croton:2005cl, Conroy:2007ik, Gallazzi:2008ku}. See \citet{Muldrew:2011go} for a comparison of 20 different environment definitions applied to a mock galaxy catalogue.

Our goal is not to review and test the various methods of quantifying environment. Our goal is to select the most isolated galaxies where environmental processes are minimized. G12 has found that low mass galaxies with $10^{7} M_{\odot} < M_{*} \leq 10^{9} M_{\odot}$ are always star-forming when located more than $1.5$~Mpc from a massive host. In our work, we assume that environment has not significantly affected such low mass galaxies and we adopt the results of G12 as our isolation criterion for low mass galaxies (also see \citet{Wheeler:2014kz}).

For low mass galaxies with $M_{*} \leq 10^{9.5} M_{\odot}$, we calculate the 2D projected distance ($d_{\rm host}$) to the nearest more massive host within a redshift of 1000\kms\ and within a 2D projected distance of 7~Mpc. Massive neighbors may be located beyond the edges of the SDSS survey, therefore we also search for massive hosts beyond the SDSS footprint using the 2MASS Extended Source Catalog \citep{Jarrett:2000fz}. A galaxy is considered massive if $M_{*} \gtrsim 2.5 \times 10^{10} M_{\odot}$ (the 2MASS magnitude $M_{K_{s}} < -23$). Therefore, massive hosts are at least $\sim 0.5$ dex more massive than the low mass galaxies. Redshifts for 2MASS hosts are obtained using a variety of other surveys and literature as described in G12. We therefore define a low mass galaxy to be isolated if $d_{\rm host} > 1.5$~Mpc. We set $d_{\rm host}~=~7~\rm~Mpc$ if no such nearby massive host is found.

Since the results of G12 apply only to low mass galaxies, the above isolation criterion is ill-defined for massive galaxies. To select a small sample of isolated massive galaxies in the NSA catalog with $M_{*} > 10^{9.5} M_{\odot}$, G12 created a modified \citet{Tinker:2011ul} group catalog where the central galaxies in the group catalog were defined to be located at least three times their halo radius from another central galaxy. We define our high mass isolated criteria by successfully recovering these isolated massive galaxies from the modified \citet{Tinker:2011ul} group catalog.

For high mass galaxies with $M_{*} > 10^{9.5} M_{\odot}$, we similarly measure the 2D projected separation to the nearest more massive galaxy by at least 0.5 dex in stellar mass within a 2D projected distance of 7 Mpc and within 1000\kms in redshift. We call this distance $d_{\rm host, 0.5}$. We place the same cut on $d_{\rm host, 0.5}$ of 1.5 Mpc as the low mass sample. 

The cut on $d_{\rm host, 0.5}$ is not applicable to the most massive galaxies in our sample because too few or no massive galaxies exist above 0.5 dex in mass. Therefore, we use the SDSS and 2MASS catalogs to measure the fifth nearest neighbor surface density ($\Sigma_{n} = n/\pi r_{n}^2$), where n is smaller if less than five neighbors are found within 7 Mpc. To be isolated, we require $\Sigma_{n} < 1 \rm~Mpc^{-2}$. This cut on $\Sigma_{n}$ is motivated by the modified \citet{Tinker:2011ul} group catalog, since we maximize the recovery of this catalog by cutting on $\Sigma_{n} < 1 \rm~Mpc^{-2}$. Decreasing the cutoff in $\Sigma_{n}$ does not appreciably effect our results. Removing this cut entirely will affect our results by including one gas-depleted galaxy at high masses relative to the rest of the sample (see \S\ \ref{subsec_m_star_m_gas}). Therefore to be considered isolated, we require a massive galaxy to have both $d_{\rm host, 0.5} > 1.5 \rm~Mpc$ and $\Sigma_{n} < 1 \rm~Mpc^{-2}$.

Our isolated low mass galaxies may have a nearby neighbor with similar stellar mass that has not been captured by $d_{\rm host}$ \citep[e.g.][]{Stierwalt:2015ju}. We calculate the 2D projected separation to the nearest, more massive galaxy ($d_{\rm pair}$) for all galaxies with $M_{*}~<~10^{10}~M_{\odot}$. We use this environment variable to avoid radio beam confusion. We only require that $d_{\rm pair} > 2r_{\rm beam}$ where the beam radius ($r_{\rm beam}$) is defined in Equation \ref{eq_r_beam} below. Both halves of the overlapping pair are removed. We discard 53 galaxies, both isolated and non-isolated, with overlapping beams, all of which are from the ALFALFA survey. We find no motivation to place further restrictions on $d_{\rm pair}$. In summary: 

\begin{enumerate}[(a)]
\item A low mass galaxy is isolated if $M_{*}~<~10^{9.5}~M_{\odot}$ and $d_{\rm host}~>~1.5~\Mpc$. 
\item A high mass galaxy is isolated if $M_{*}~>~10^{9.5}~M_{\odot}$, $d_{\rm host, 0.5}~>~1.5~\Mpc$ and $\Sigma_{n}~<~1~ \rm Mpc^{-2}$. 
\end{enumerate}

\noindent We refer to all galaxies that do not meet our isolation criteria as ``non-isolated" for the duration of this paper. 

In terms of central and satellite galaxies: All isolated galaxies are central galaxies. All satellite galaxies are categorized as non-isolated. Due to the strictness of our isolation criteria, many non-isolated galaxies would be categorized as central galaxies. We define \NISOINNSA~galaxies as isolated and \NNONISOINNSA~galaxies as non-isolated in our catalog of \NALLINNSA~total galaxies. We generally assume that our isolated sample is largely comprised of disk galaxies, even at the lowest masses.

\subsection{HI Observations} 
\label{subsec_hi_obs}

We have obtained \NTOTALHIGHQUALITY~HI observations with S/N greater than 2 in an observing program spanning 2005 July to 2014 Sept, see G06 for an introduction to these data. After 2008, we focus our observations on isolated, low mass, highly inclined (an axis ratio less than 0.65) galaxies from the NSA catalog described above with $d_{\rm host} > 1.5 \rm ~Mpc$, $M_{*} \leq 10^{9.5} M_{\odot}$.
 
All single-dish, unresolved HI measurements were obtained using the 305~m Arecibo Telescope (AO) and the 100~m Green Bank Telescope (GBT). For observations at AO in 2005 (hereafter AO2005), we observed with the L-Band Wide receiver using 1024 resolution channels with a bandwidth of 12.5~MHz and a velocity resolution of 2.6\kms. For observations at AO in Spring 2013, Spring 2014 and Fall 2014 (hereafter AO2013, AO2014a and AO2014b), we observed using the L-Band Wide receiver using 4096 resolution channels with a bandwidth of 12.5~MHz and a velocity resolution of 0.65\kms. For observations at GBT in 2005, 2006 and 2008 (we group all observations together as GBT2008), we observed using the L-band receiver using 8192 channels with a bandwidth of 12.5~MHz and a velocity resolution of 0.32\kms. In all cases, position switched ON/OFF observing and ON/OFF noise-diode calibration was used.

Our target galaxies have an average Petrosian 90\% light radius ($r_{90}$) of $\sim20''$, significantly smaller than the GBT and AO beam sizes ($9'$ and $3.5'$, respectively), therefore flux attenuation and pointing offsets should not significantly affect our observations. We remove galaxies from our sample in the rare instances where the beam size is less than $r_{90}$. We detail each observing run in Table \ref{tbl_obs}, including average integration times, the number of reduced spectra, the number of high S/N spectra, the number of non-detections and the number of unstable spectral baselines. For \NTOTALREDUCED~successfully reduced spectra, we make \NTOTALHIGHQUALITY~detections of galactic HI emission. We do not detect HI emission in \NTOTALNONDETECT~cases, most of which are non-isolated galaxies from GBT2008. We discuss our non-detections further in \S\ \ref{subsec_non_detect}. 

\subsection{HI Data Reduction and Measurements}
\label{subsec_hi_measurement}

While there exist many methods for determining the HI properties of 21-cm emission lines, we implement a simple and consistent method optimized to minimize user intervention and maximize reproducibility of our measurements. We measure the shape and total emission of each single-dish HI observation without assuming a functional form for the 21-cm line. We also automate our entire procedure following a preliminary quality-control check. Automating our measurement procedure enables us to use a Monte Carlo method for estimating random errors in our HI parameters (see \S\ \ref{subsec_hi_error}).

We first use our observing notes to identify \NTOTALBADBASELINES~spectra with bad spectral baselines which can significantly affect our HI measurements. If the signal cannot be differentiated from a noisy baseline then we discard the observation and add the galaxy to a future observing queue. Next, we perform a linear baseline fit to each polarization mode outside of the region of each galaxy's HI emission at the SDSS systemic velocity. Since we generally target isolated galaxies, we do not find overlapping signals in our spectra. We subtract the baseline fits from each polarization mode and average the two polarization modes. 

To achieve higher S/N and ensure reliable detections with our measurement algorithm, we determine an optimal smoothing length for each observing resolution that produces the most accurate HI parameters in our reduction pipeline. We test for optimal smoothing lengths by degrading synthetic Gaussian signals over a range of realistic widths, integrated fluxes and signal-to-noise levels. We use Hanning windows to smooth spectra to a resolution of 5\kms for all of our observations. 

For each spectrum, we identify the systemic velocity channel using the optical SDSS redshift velocity. If a spectrum does not have a signal-to-noise greater than 2 within 100\kms of the systemic velocity, we flag the observation as a non-detection. We discuss non-detections as well as the sensitivity of our observations in \S\ \ref{subsec_non_detect}.

Next we identify the peak flux, $f_{\rm peak}$, within a 500\kms spectral window around the systemic velocity. We expand outward from the peak flux velocity channel and set the boundaries of the emission to the first locations where the spectrum drops below zero flux. We locate these boundaries using a copy of the spectrum that has been smoothed by twice the Hanning window smoothing length. This extra smoothing eliminates any spurious noise features within the HI emission that may drop below zero and cut the emission profile artificially.

We measure the velocity width at $0.5f_{\rm peak}$, $W_{50}$, by measuring the distance in velocity space between the first and last spectral channels with flux above $0.5f_{\rm peak}$. To ensure we have measured the most accurate value of $W_{50}$, we interpolate between the neighboring spectral channels on either side of the channel closest to $0.5f_{\rm peak}$. 

We measure $W_{50}$ in order to directly compare with published ALFALFA results. However, since we have no reason to assume that $W_{50}$ is the best probe of the maximum circular velocity of all disk galaxies, we also measure $W_{20}$, the velocity width at $0.2f_{\rm peak}$. While $W_{20}$ is more sensitive to noise in the wings of the HI profile \citep{Koribalski:2004cv}, our high S/N isolated galaxies may offer us a deeper dynamical probe into the HI disks than $W_{50}$ and $W_{20}$. We therefore focus on $W_{20}$ in our analysis below. We note that $W_{20}$ is typically 25\kms larger than $W_{50}$ in our isolated sample, similar to \citet{Koribalski:2004cv}.

We compute the integrated flux density, $S_{21}$, by integrating the flux between the channels of the emission line profile boundaries. In order to remove excess flux due to noise in the wings of the profiles, we fit a second order polynomial to the minima on either side of the emission profile between $0.5f_{\rm peak}$ and where the flux goes to zero, similar to the algorithm of \citet{Springob:2005db}. Using these fits, we remove spurious signal from the wings of the profile that tends to positively bias our flux measurements. We have tested this method by degrading high S/N spectra and then measuring the flux and HI profile widths as the S/N decreases. We found this method effective down to S/N of 2.

We calculate the root mean square (rms) of the noise of each spectrum, $\sigma_{\rm rms}$, by identifying the largest signal- and RFI-free spectral chunks on either side of the HI emission profile with minimum window widths of 200\kms. Two windows are selected on either side of the emission line, beginning $\pm W_{50}/2$ away from the profile edges and extending out to the maximum extent of the spectral range. Each window range is then optimized by shrinking the window until the mean flux is less than 0.5 mJy or the minimum window width is achieved. The restriction on the mean flux ensures that we select a chunk of the spectrum where the baseline is stable and no significant emission is contaminating our noise measurement. We track the mean flux in the signal-free and RFI-free portion of the spectrum to monitor baseline stability. We eliminate RFI by masking any single channels with 6-sigma emission or greater. We calculate the rms as the root mean square of the two signal-free, RFI-free and baseline stable spectral chunks. 

If multiple observations exist for a galaxy at the same telescope and observing run, each rms weighted spectrum is co-added together. For multiple observations of the same galaxy over different observing runs, we compute the error weighted mean of each HI parameter measured with our reduction method. We tune and test our HI measurement method with the \ALFALFA~data in \S\ \ref{subsec_alfalfa}. We calculate a S/N measurement for all of our observations following the ALFALFA survey, adopted from Equation 2 of \citet{Haynes:2011en}, 

\begin{equation}
S/N = \left( \frac{S_{21}}{W_{50}} \right) \frac{w_{\rm smo}^{1/2}}{\sigma_{\rm rms}},
\label{eq_sn}
\end{equation}

\noindent with each variable as defined above and where $w_{\rm smo}~=~W_{50}/10$[km/s]. Hanning smoothing redistributes spectral noise and therefore increases the signal-to-noise ratio of our spectra. This smoothing creates dependancies in each channel on neighboring channels. This dependance is accounted for in $w_{\rm smo}$. \citet{Haynes:2011en} calculates $w_{\rm smo}$ as half of the number channels across the smoothed 50\% HI line width. So $w_{\rm smo}$ represents the number of independent channels in each HI line.
\subsection{HI Parameter Error Estimation} 
\label{subsec_hi_error}

To estimate random errors in our HI parameter measurements, we apply a Monte Carlo (MC) technique. For each MC iteration, we add a random noise realization to the fiducial observed signal using our estimate of the rms noise in each spectrum and then remeasure the HI parameters over 1000 trials. The errors of each HI parameter estimate are the rms of the difference between each MC realization and the fiducial value of the parameter estimate.

We note that our method of error estimation does not consider baseline stability and its influence on the integrated flux density, which may be a source of significant uncertainty. It is for this reason that we have visually inspected and removed any spectra with obvious baseline instability. We have also made an effort to obtain multiple observations of the same galaxies to inspect any additional systematic error (see the last paragraph of \S\ \ref{subsec_alfalfa}). 

\begin{figure}[t]
\epsscale{1.2}
\plotone{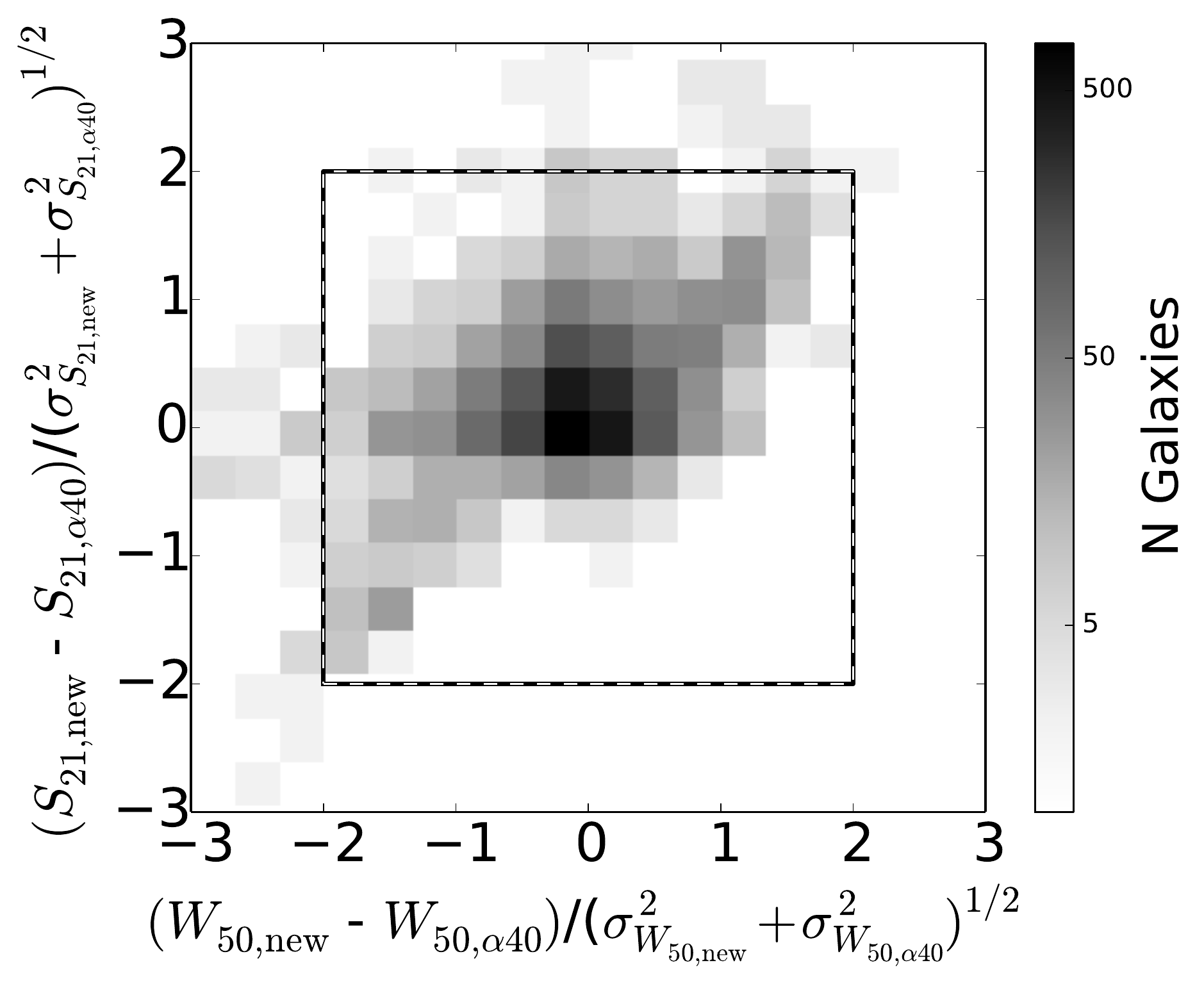}
\caption{Two dimensional distribution of the difference between our re-measurement of the ALFALFA spectra and the original ALFALFA measurement in units of $\sigma$, where $\sigma$ is the error for each galaxy from the two data sets added in quadrature. The difference in 50\% velocity width $W_{50}$ and HI flux density $S_{21}$ are plotted on the x- and y-axis respectively. The number of galaxies in each bin are shown in the greyscale bar on the right. The over plotted box represents our selection boundaries ($\pm 2\sigma$) on the data. Any measurement outside of the box is discarded as a failed measurement. We discuss the asymmetry in the distribution at negative differences in $W_{50}$ and positive differences in $S_{21}$ which is caused by a difference in measurement methods in \S\ \ref{subsec_alfalfa}.}
\label{fig_alfalfa_compare}
\end{figure}

\subsection{Homogeneous HI Measurements of ALFALFA}
\label{subsec_alfalfa}

In addition to our observations of isolated low mass galaxies, we have re-measured all Arecibo Legacy Fast ALFA spectra overlapping with our NSA catalog (from the 40\% data release of \citet{Haynes:2011en}). The ALFALFA survey is a blind, single-dish, flux-limited, HI survey. The \ALFALFA~data release covers $\sim$2800 deg$^2$ of the sky and provides a significant amount of overlap with the SDSS footprint. We match ALFALFA HI sources where the HI center is within $25''$ of the optical center of our NSA catalog, given the pointing accuracy of the survey \citep{Haynes:2011en}. We use the ALFALFA data to compliment our isolated low mass sample with both non-isolated galaxies and massive galaxies.
 
Re-measurements are preformed primarily to maintain a homogeneously measured data set and to apply our MC error calculation. We do not visually inspect or baseline fit the \ALFALFA~data, instead we begin with the baseline-corrected spectra from ALFALFA and cut on their quality code of 1 to ensure reliable detections. We run these \ALFALFA~data through our HI parameter measurement steps described in \S\ \ref{subsec_hi_measurement}, convolving a 3 channel Hanning smoothing window with the downloaded \ALFALFA~spectra. 

Figure \ref{fig_alfalfa_compare} shows the difference between our re-measurements and the original \ALFALFA~measurements in units of the error of each HI parameter between the two data sets added in quadrature. In general, both our remeasurements of $W_{50}$ and $S_{21}$ agree extremely well with the \ALFALFA~measurements. As expected, we find that just under 5\% of galaxies lie beyond the $\pm 2 \sigma$ level (\NHIALFALFA~HI observations in Figure \ref{fig_alfalfa_compare} (\NHIALFALFACUT~outliers lie beyond the \NALFALFARCUT-$\sigma$ boundaries in either $W_{50}$ or $S_{21}$). We remove galaxies beyond the $2 \sigma$ window from our sample because their emission line boundaries are ambiguous and we require minimal scatter in our analysis below. We also discard any galaxy where the error on either $W_{50}$ or $M_{\rm HI}$ is two times the measured value or more. We recover \ALFALFATWOSIGPERC\% of the complete \ALFALFA~dataset with our remeasurements.

We note an asymmetry in the 2D distribution of differences in $W_{50}$ and $S_{21}$ in Figure \ref{fig_alfalfa_compare} where we underestimate $W_{50}$ and overestimate $S_{21}$ relative to ALFALFA. The cause of this is due to a difference in measurement methods. Since we measure $W_{50}$ at 50\% the peak flux without accounting for noise or the shape of the HI profile, our value of $W_{50}$ will always be larger than ALFALFA since they account for the shape of the HI profile when calculating such line-widths. This will tend to decrease the value of the peak flux and therefore ALFALFA will always measure an equal or larger value of $W_{50}$. The spectral boundaries for the ALFALFA measurement are manually marked by eye (see \citet{Haynes:2011en}, \S\ 3 item 6). It is possible that this tends to remove some of the actual signal that overlaps with the noise features in the wings of the HI profile while our method of automatically marking spectral boundaries tends to include some fraction of the noise in the wings of our profiles. Since less than 1\% of our measurements are beyond $\pm 1\sigma$ in this region, and 3\% are beyond $\pm 0.5\sigma$, we proceed with our measurements. We also note that our analysis proceeds with $W_{20}$, which is less sensitive to noise in the peak flux and a more effective measurement of line-width in the lowest mass galaxies. We will discuss this further in our future paper on the baryonic Tully-Fisher (BTF) relation.

\begin{figure}[t]
\epsscale{1.19}
\plotone{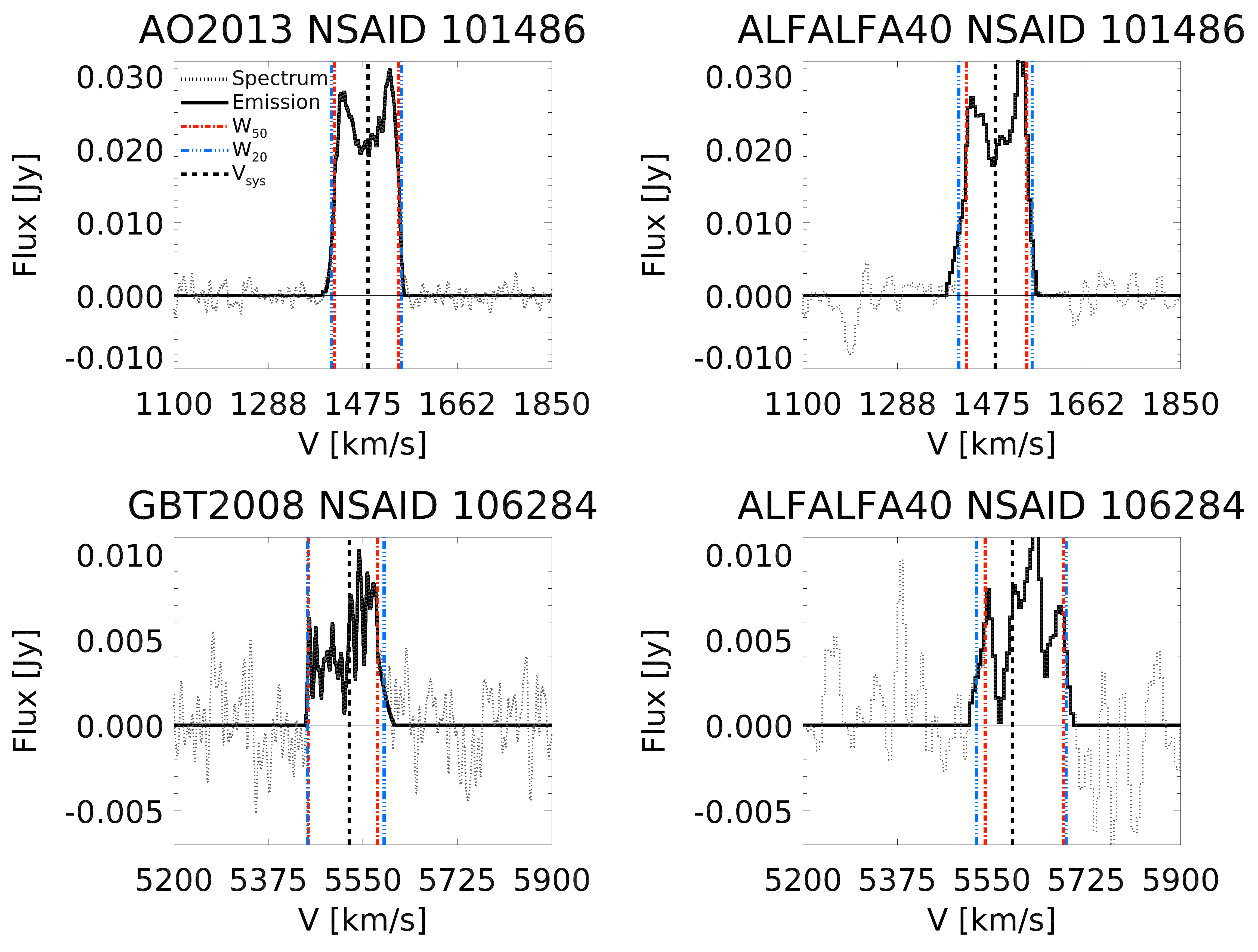}
\caption{A sample pair of overlapping HI spectra between our sample and \ALFALFA. The left spectra are from our observations. The right spectra are from the \ALFALFA~data set. In all panels, the smoothed spectrum is plotted as a grey dotted line, the emission region is identified in solid black, the 50\% and 20\% flux boundaries are identified with red dash-dot and blue dash-dot-dot vertical lines respectively. The central velocity is marked with a vertical dotted black line. The top spectra are representative of a straight-forward measurement, while the bottom spectra are more ambiguous due to the difference in spectral resolution. The more ambiguous cases will result in larger errors in $W_{50}$ and $S_{21}$ due to our method of estimating errors.}
\label{fig_sample_obs}
\end{figure}

Our HI observations overlap with \NAOverlap~of the \ALFALFA~sample. We compare our HI parameter estimates to the original \ALFALFA~catalog for each overlapping galaxy. Most of our HI parameter estimates agree when compared to the original values from the \ALFALFA~data release. However, several measurements of $S_{21}$ disagree between our GBT2008 observations and the original \ALFALFA~catalog. We assume that both our measurements and the \ALFALFA~measurements are correct and we attribute any differences to systematic errors that we have underestimated with our error estimation method (most likely due to baseline instability in the GBT2008 data). In order to bring our GBT2008 measurements into agreement with the ALFALFA catalog, we add 0.2 Jy km/s in quadrature to all of the GBT2008 $S_{21}$ uncertainties. We present two pairs of overlapping galaxy spectra in Figure \ref{fig_sample_obs} for reference. In this figure, the top pair is a straightforward comparison while the bottom pair is more ambiguous.

\subsection{Non-Detections}
\label{subsec_non_detect}

Our observing strategy using the AO is to integrate until we measure a significant HI signal or until we reach a 25 minute on-source integration time when possible. Most of our observations result in detection, but in some cases we do not detect significant HI emission. We define a non-detection as an observation with S/N~$< 2$, the threshold where our measurement method begins to fail. Out of our \NTOTALREDUCED~reduced spectra, \NTOTALNONDETECT~have a S/N less than 2. Nearly all of these low S/N observations are from GBT2008 since these observations were made before we implemented a fixed observing time criteria and focused our observations on isolated galaxies. Given the smaller dish size, the average GBT exposure leads to smaller S/N than an equivalent AO integration. \textit{Out of \NTOTALISOLATED~isolated galaxies we observed using our post-2008 observing strategy, \NTOTALISOND~are non-detections}.

For all HI non-detections, we determine upper limits on $S_{21}$. First, we make preliminary estimates of atomic gas masses using the stellar-to-atomic gas mass relation in \S\ \ref{subsec_m_star_m_gas}. Next, we estimate the inclination corrected 50\% velocity widths using the relation between baryonic mass and 50\% inclination-corrected velocity widths, similar to the relation we describe in \S\ \ref{subsubsec_mv_relation}. We inclination de-correct this 50\% velocity width to obtain an estimate of $W_{50}$. We initialize the value of $S_{21}$ assuming a gas fraction of $f_{\rm gas} = 0.98$. 

Next, we create a synthetic Gaussian 21-cm emission line with a FWHM equal to $W_{50}$ and a total integrated flux of $S_{21}$. We inject this initial Gaussian signal into 100 noise realizations using the measured $\sigma_{\rm rms}$ of our observation. After each noise realization, we reduce $S_{21}$ by 10\% and repeat a new set of noise realizations while we keep the FWHM of the Gaussian fixed. We repeat this procedure until the S/N drops below 2. The final injected values of $S_{21}$ are reported as the upper limits of each non-detection. These upper limit estimates are plotted as light-colored, downward pointing arrows in all plots. Upper limits are excluded from our velocity-mass-size scaling relations and baryon fraction analysis since their velocity widths are contrived.

\ALFALFA~code 1 sources are all above the ALFALFA survey's S/N threshold of 6.5, therefore upper limit calculations are unnecessary. Code 2 sources from \ALFALFA~have lower S/N and since they do not significantly impact our results below, we do not consider these data in our analysis.
\subsection{Calculated and Inferred Quantities}
\label{subsec_quant}

We next describe the quantities derived directly from our HI line emission parameters and the optical parameters from the NSA catalog. In Table \ref{tbl_hienv}, we present a small sample of isolated low mass galaxies from both our \NHIOurs~original observations and the re-reduction of \NHIALFALFAHIQ~\ALFALFA~observations. Our complete data set is available upon request and will be publicly released once our HI observation program has concluded.

Distances ($D$) reported here are luminosity distances calculated from peculiar motion corrected redshifts \citep{Willick:1997gg}. For our calculations and analysis, we assume that all of our galaxies are oblate, axisymmetric spheroids with symmetric mass distributions. We infer the observed inclination, $\rm sin ~ \it i$, from the axis ratio as,

\begin{equation}
\rm sin \it ~i = \sqrt{\frac{1 - (b/a)^2}{1 - q_0^2}},
\end{equation}

\noindent where $q_0$ is the intrinsic axis ratio, $b/a$ is the observed axis ratio and $i$ is the inclination angle. Note that we limit the maximum value of the axis ratio $b/a$ to 0.9995 to avoid dividing by zero in Equation \ref{eq_v_50i}. 

The value of $q_0$ is traditionally fixed at about 0.2 at all stellar masses \citep[e.g., G06, ][]{Begum:2008cx}. See \citet{Yuan:2004kr} for a brief discussion on the history of $q_0$ as well as the distribution of $q_0$ as a function of morphology. The effect of $q_0$ on inclination corrections can be significant enough to change the slope and possibly affect the scatter of the BTF relation, \citep[see][Figure 1.8]{Singhal:2008tp}. We plan to discuss the effect of $q_0$ further in a future paper, where we will analyze circular velocities using the traditional value of $q_0 = 0.2$ as well as with a prescription for a changing $q_0$ as a function of stellar mass. In this work, we set $q_0 = 0.2$ throughout our analysis below.

We remove redshift broadening and we inclination correct the HI line-widths to estimate a measure of maximum rotational velocity of the stars and HI gas disk in each galaxy as,

\begin{equation}
V_{X,i} = \frac{W_{X}}{2 \rm sin \it ~i(1 + z)},
\label{eq_v_50i}
\end{equation}

\noindent with $X = 50$ or $X = 20$ for either the 50\% or 20\% 21-cm line-width, respectively. We perform these inclination corrections to $W_{X}$ but we do not correct for non-circular motions (i.e. turbulence) which is typically assumed to be small (less than 20 \kms). Note that in previous work, G06 corrected for non-circular motions. We note that the 20\% velocity width has been shown by some authors to provide a better estimate of the true maximum circular velocity of the galactic disk \citep{Verheijen:1997vb, AvilaReese:2008gz}. We plan to discuss the effect of non-circular motions on HI velocity widths, as well as the accuracy of HI line-widths as a probe of maximum circular velocity in a future paper. For now, we assume $V_{W20,i}$ is an accurate representation of maximum circular velocity.

We find that the inclination measurement for a small fraction of galaxies in our catalog fail due to the presence of central galactic bars. Inclination measurements fail when a face-on galaxy resembles the light distribution of a more edge-on galaxy due to the thin, centrally concentrated distribution of light of a galactic bar. We remove any barred galaxies that have been identified by the Galaxy Zoo project \citep{Hoyle:2011bk} from our analysis of circular velocities. We preserve these barred galaxies unless specified in our analysis below. 

\begin{figure}[t]
\epsscale{1.2}
\plotone{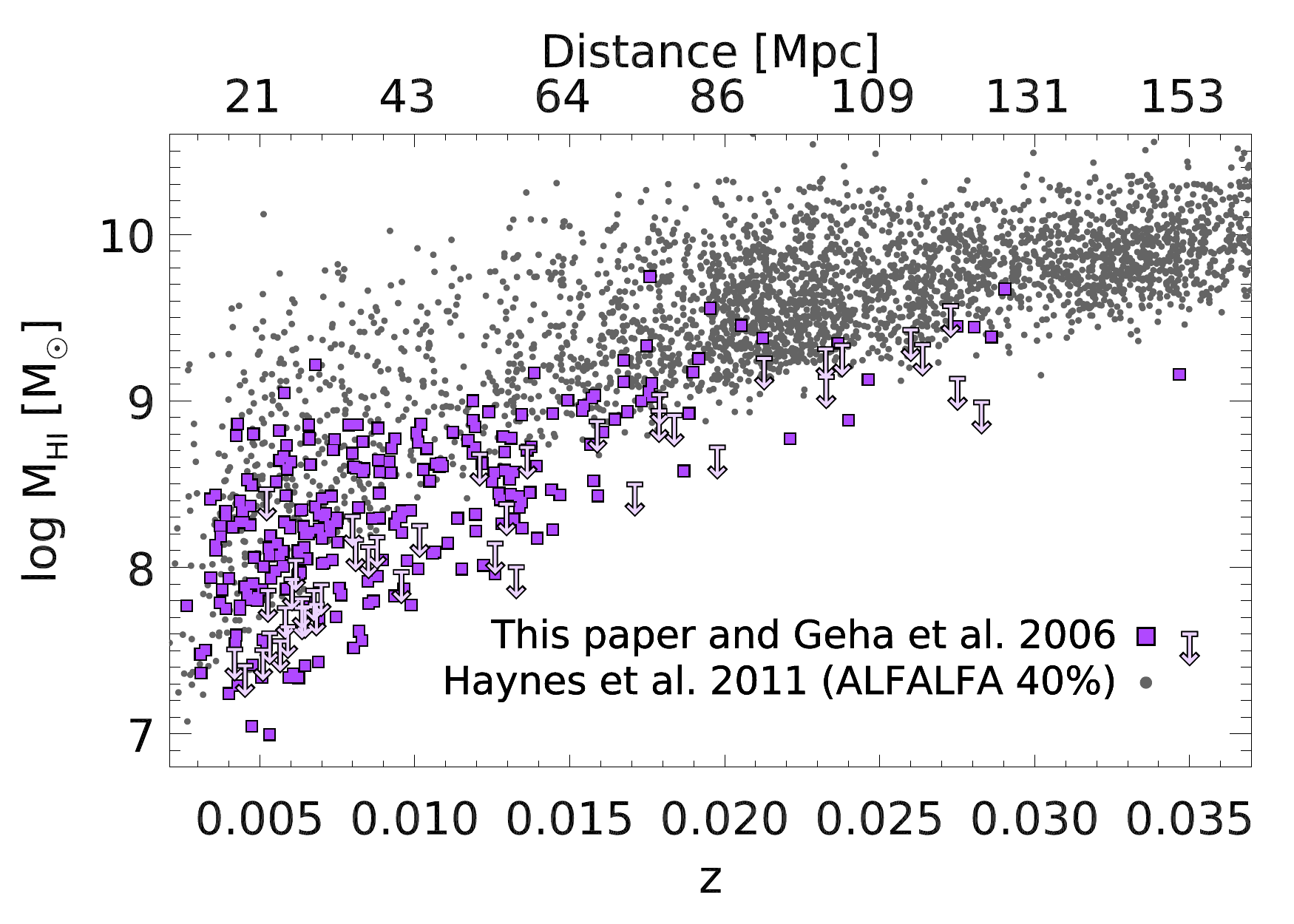}
\caption{HI masses versus peculiar motion corrected redshifts/distances of NASA Sloan Atlas galaxies that have been observed in 21-cm emission. Grey data points are our ALFALFA re-measurements. Purple squares are our own new and previously published (G06) observations with downward pointing, light-colored arrows representing upper limits on HI masses for non-detections. Our HI observations complement the ALFALFA survey at low masses and also overlap the ALFALFA sample in several cases as a consistency check between datasets.}
\label{fig_z_hi_mass}
\end{figure}

We calculate the HI mass from the $S_{21}$ flux using the standard formula \citep[e.g.][]{Haynes:1984el} as,

\begin{equation}
M_{\rm HI} = 2.356\times10^5 \left(\frac{D}{\rm Mpc}\right)^2 \frac{S_{21}}{\rm Jy~km~s^{-1}}~[M_{\odot}],
\end{equation}

\noindent which assumes the HI gas disk is optically thin. We do not correct for HI self-absorption so our 21-cm flux measurements are lower limits due to the self-absorption of HI flux in a galaxy \citep{Haynes:1984el, Haynes:2011en}.

We present HI masses as a function of redshift in Figure \ref{fig_z_hi_mass}. This figure shows the \ALFALFA~dataset as grey points with our data and data from G06 plotted in purple. Our observations probe to lower HI masses at all redshifts than \ALFALFA. 

We calculate atomic gas masses as $M_{\rm gas}~=~1.4 M_{\rm HI}$ where the factor of 1.4 corrects for helium assuming solar abundances. We calculate ``cold" baryonic masses as $M_{\rm baryon}~=~M_{\rm gas}~+~M_{*}$ with the atomic gas mass fraction defined as,

\begin{equation}
f_{\rm gas}~=\frac{M_{\rm gas}}{M_{\rm baryon}}.
\end{equation}

\noindent Note that our final cold baryon masses do not include dust, molecular gas or ionized gas, the combination of which may contribute significantly to the baryon content of our galaxies \citep{McGaugh:1997fk, Cortese:2012cv, Gnedin:2012ip}. Further quantities, including stellar mass surface density and baryon fraction are described in the appropriate sections below. We discuss the impact of molecular gas on our results in section \S\ \ref{subsec_gfhs}.

\subsection{HI Detection Limits, Isolated vs Non-Isolated Sample Sizes and Surface Brightness Limits}
\label{subsec_sensitivity}

Here we discuss the relative sample sizes of our isolated and non-isolated galaxies as well as the surface brightness limitations of the NSA catalog. We also discuss how a flux-limited HI survey affects our results and subsequent conclusions.

\subsubsection{HI Detection Limits} The minimum HI mass detected by ALFALFA that overlaps with our catalog is $10^{\ALFALFAMinHIMass} M_{\odot}$ while the minimum HI mass detected by our observations is $10^{\OurMinHIMass} M_{\odot}$. Since the ALFALFA survey is flux-limited, it has not detected low mass, HI-depleted galaxies such as satellites in dense groups and clusters (see \S\, 6 of \citet[][]{Haynes:2011en} for a thorough discussion of the \ALFALFA~completeness). For the duration of this paper, we define detection limits as the minimum HI mass of either our data set or the \ALFALFA~catalog that overlaps with the NSA catalog.

When the ALFALFA survey is combined with our observations, the HI sample we study here is largely incomplete and biased towards blue, HI-rich, late-type galaxies \citep{Haynes:2011en, Huang:2012gk}.The median $g - r$ color for the ALFALFA survey and our HI data is \ALFALFAMedianColor~and \OurMedianColor, respectively. Satellite galaxies in the local group \citep{Spekkens:2014he} and in the Virgo cluster \citep{Cortese:2011hz} are observed to be severely HI-depleted and would not be detected by either our observations or by ALFALFA. Therefore, even the relatively HI-depleted galaxies in our sample are still quite HI-rich compared to the average red sequence galaxy. 

Due to our HI observation strategy, we have detected nearly every isolated galaxies identified optically. For the few isolated galaxies that we have not made a detection, we calculate upper limits at relatively large atomic gas fractions. This is one of the primary results of our paper: \textit{all of the isolated galaxies that we have observed have either been measured at large atomic gas fractions or large atomic gas fractions cannot be ruled out via their upper limits} (see \S\,\ref{subsec_m_star_f_gas}). As for our comparison to non-isolated galaxies, our sample only glimpses into the HI-poor population that we expect to find in satellite galaxies. In our analysis below, we discuss these points further in context with our results. 

\subsubsection{Isolated versus non-isolated galaxy sample sizes} In order to disentangle internal galaxy evolutionary processes from the effects of environment, we select only the most isolated galaxies in the NSA catalog given our definition in \S\, \ref{subsec_env}. Despite our overall HI sample being largely incomplete, we have made every effort to observe similar numbers of isolated and non-isolated galaxies at low stellar masses. Our analysis therefore assumes that the outliers we measure are significant and not simply an effect of the relative sample sizes between isolated and non-isolated galaxies.

For galaxies with stellar mass below $10^{\LowMassThreshWithGas} M_{\odot}$, we have \NumberLowMassSatsWithGas~non-isolated galaxies and \NumberLowMassIsoWithGas~isolated galaxies with HI observations. We achieve an average fraction of isolated galaxies with HI observations of \IsolatedFractionWithGas~for $M_* < 10^{\LowMassThreshWithGas} M_{\odot}$, while the average fraction of isolated galaxies in the entire NSA catalog is \FRACISONSA. We assume that this relative sample size should allow us to rigorously compare the relations, distributions and outliers as a function of environment.

\subsubsection{Surface brightness limitations} Even though the NSA catalog has been optimized for nearby galaxies in the SDSS, the catalog still suffers from low surface brightness completeness limitations. At a half-light surface brightness $\mu_{50,r} \sim 23.5 \rm~mag~arcsec^{-2}$, the SDSS completeness drops below 50\%. At $\mu_{50,r} \sim 24.0 \rm~mag~arcsec^{-2}$ the SDSS completeness drops below 10\% \citep[][G12]{Blanton:2005gz, Blanton:2008il}.

This surface brightness incompleteness impacts our study if either of two extreme situations exist: a.) we preferentially miss HI-poor galaxies or b.) we preferentially miss gas-rich galaxies due to incompleteness.

If we preferentially miss HI-poor galaxies, then a HI-poor population of isolated galaxies might exist that has gone undetected. G12 compared the surface brightness distributions of low mass non-isolated galaxies to isolated galaxies via the two-sided K-S test and found that the two samples are likely drawn from the same distribution. Therefore, we have little evidence that suggests a larger number of HI-poor isolated galaxies at the low surface brightness end of the distribution are missing from our sample compared to the number of HI-poor non-isolated galaxies.

The recent discovery of ultra diffuse galaxies (UDGs) in the Coma cluster by \citep{vanDokkum:2015ks, vanDokkum:2015vc} using the Dragonfly Telephoto array \citep{Abraham:2014dg, Merritt:2014ha} suggests that a significant population of galaxies with extremely low central surface brightness exists in galaxy clusters and the local field ($24 < \mu_{g,0} < 26 \rm~mag~arcsec^{-2}$) \citep[][]{Dalcanton:1997kf}. This may indicate that more low mass galaxies (with $-13 < M_V < -17$) are missing from our SDSS sample than previously thought. It is possible that a similar population of UDGs exist in ``the field". However, we have no evidence at this time that we are preferentially missing a \textit{significant} number of isolated HI-poor UDGs.

We have no evidence that a significant number of HI-rich LSB galaxies are missing from our sample due to the following logic: We find that galaxies with surface brightnesses below $\mu_{r, 50} < 23 \rm~mag~arcsec^{-2}$ have a median $f_{\rm gas}$ of $0.7\pm0.2$ (also see \citet{Schombert:2001gc}). Therefore, at all stellar masses where $\mu_{r, 50} < 23 \rm~mag~arcsec^{-2}$, galaxies should have atomic gas masses equal to or significantly greater than their stellar masses. These sources would be detected by ALFALFA but would have no optical counterparts in SDSS. The \ALFALFA~data release has found just $\sim 50$ extragalactic HI sources (out of 15,855) without optical counterparts in the SDSS DR7 and associated catalogs \citep{Haynes:2011en, Cannon:2015cu}. This suggests that while such a population exists, these ``dark galaxies" are rare.
\section{Results}
\label{sec_results}
\begin{figure*}[t!]
\epsscale{1.22}
\plotone{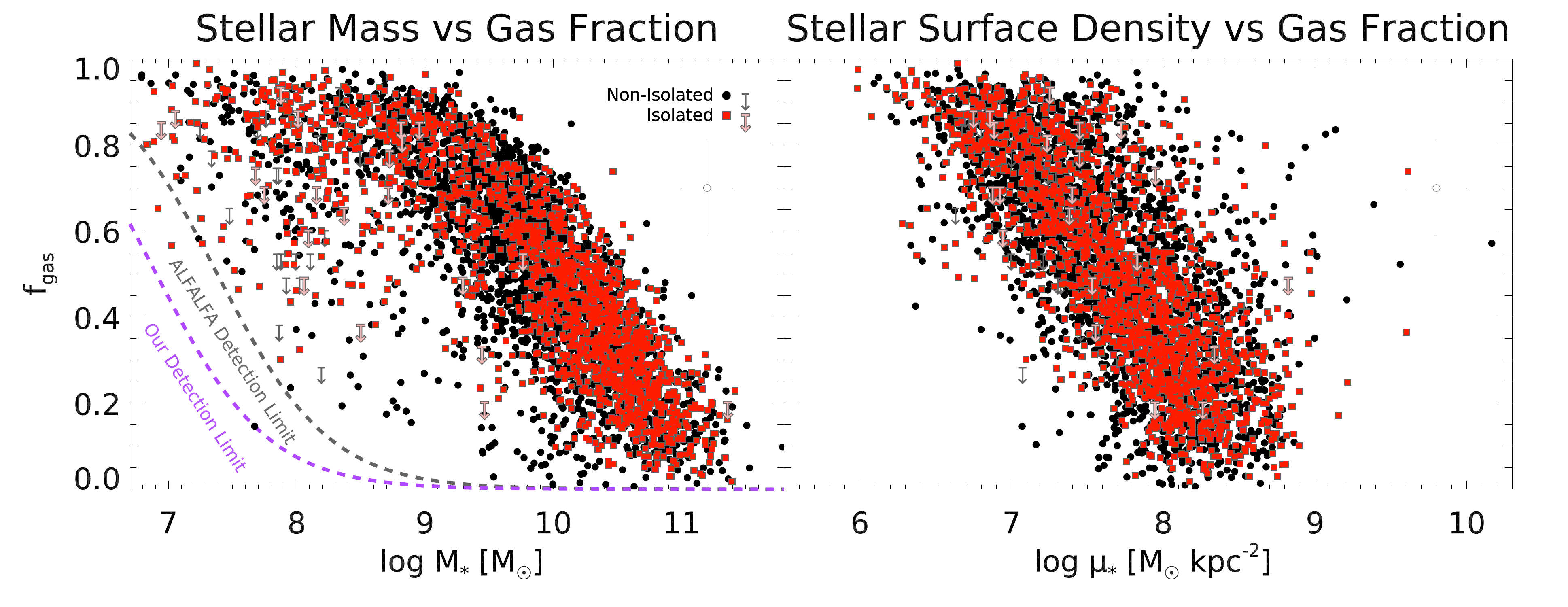}
\caption{(Left) Atomic gas fraction versus Stellar mass. Black dots are galaxies that have not passed the isolation criteria described in \S\ \ref{subsec_env}. Red squares have passed the isolation criteria for each mass regime described in \S\ \ref{subsec_env}. We plot the median error bar for the entire sample as the open point, which is representative of the rest of the sample. This relation saturates at about $10^9 M_{\odot}$ in stellar mass. The dashed grey line represents the detection limit (minimum HI mass detected) of the overlap of our sample with the ALFALFA survey ($10^{7.2} M_{\odot}$). The dashed purple line represents the detection limit (minimum HI mass detected) of our HI observations ($10^{\OurMinHIMass} M_{\odot}$). Non-isolated galaxies are scattered to lower $f_{\rm gas}$ compared to isolated galaxies at nearly all stellar masses, indicating that environment affects the atomic gas fraction of non-isolated galaxies. Low mass isolated galaxies are not found below an $f_{\rm gas}$ of 0.3. (Right) Atomic gas fraction versus stellar mass surface density. (Right) We divide stellar mass by the square of the effective radius to show that stellar mass surface density is a better predictor of atomic gas fraction than stellar mass alone.}
\label{fig_m_star_f_gas}
\end{figure*}

We begin our analysis by inspecting the atomic gas fraction of galaxies as a function stellar mass and stellar mass surface density. We then compare the relation and scatter of stellar mass to atomic gas mass in isolated environments compared to non-isolated galaxies. We analyze the velocity-mass-size galaxy relations and the impact of environment on the slope, scatter and correlation of these scaling relations. Finally, we investigate baryon fractions in these systems.

In Figures \ref{fig_m_star_f_gas} through \ref{fig_baryon_fraction}, we represent isolated galaxies with red squares and non-isolated galaxies with black dots. Whenever we exclude galaxies from our analysis, we preserve these data as grey points. We plot upper limits as light-colored, downward pointing arrows, color coded with environment. We also plot a hollow data point with a representative set of error bars wherever error bars are not explicitly presented. In Figures \ref{fig_v_r_m} and \ref{fig_baryon_fraction}, we plot non-isolated galaxies we determine to be gas-depleted as open blue dots. We have applied the Bayesian linear regression fitting method of \citet{Kelly:2007bv} for all linear fits below. This fitting method is particularly useful for linear models that also measure the intrinsic scatter of galaxy scaling relations.


\subsection{Atomic Gas Fraction Versus Stellar Mass}
\label{subsec_m_star_f_gas}

\begin{figure*}[t!]
\epsscale{1.24}
\plotone{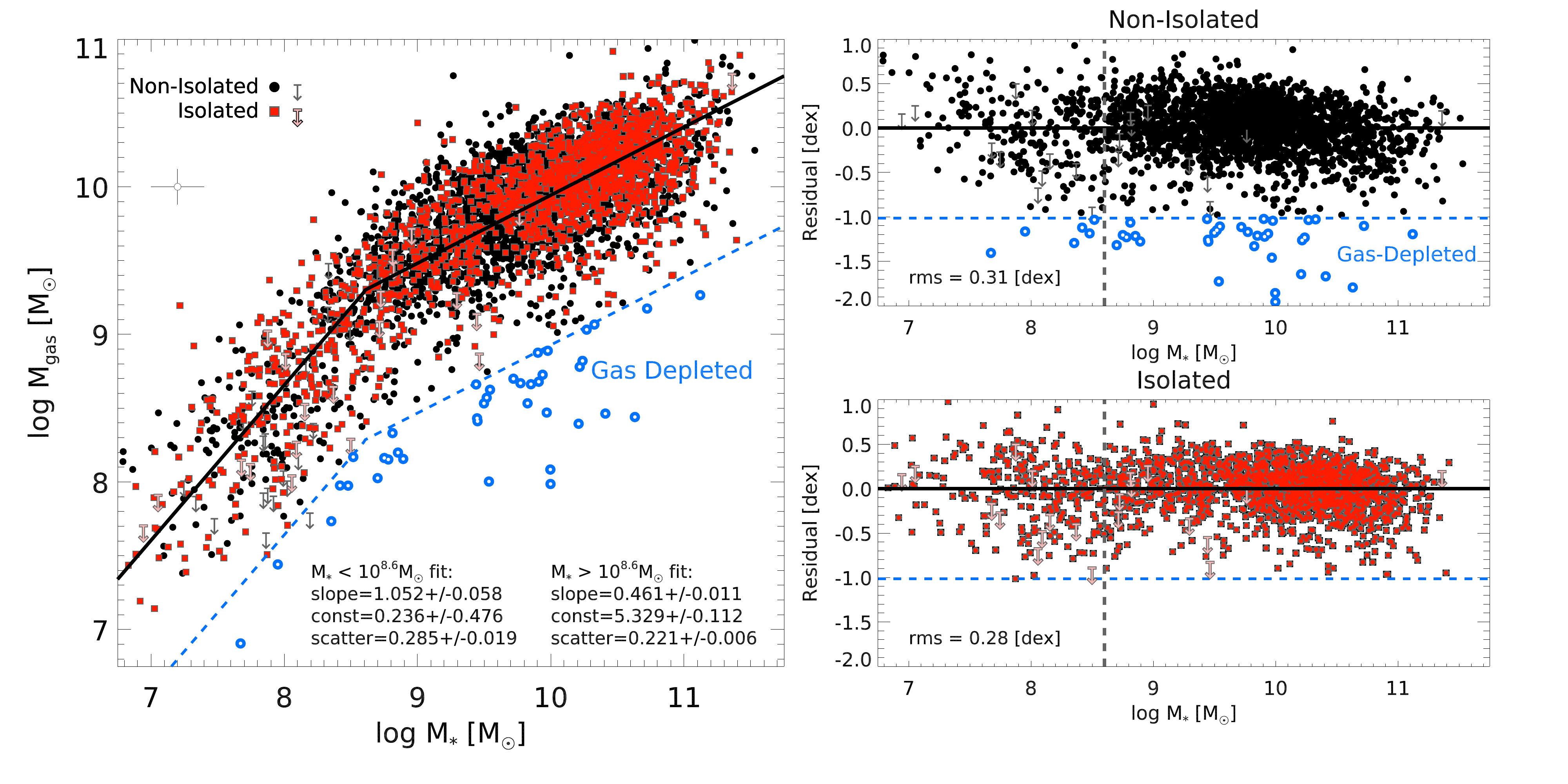}
\caption{(Left) Atomic gas mass plotted as a function of stellar mass. Black dots are non-isolated galaxies, red squares are isolated galaxies. We plot the median error bar for the entire sample as an open point, which is representative of the rest of the sample. The fits here are listed below and in Table \ref{tbl_scaling_fits} and are plotted as the solid black line. We plot a dashed blue line representing the $-1$~dex threshold where we define galaxies to be gas-depleted when located below this threshold. We identify non-isolated gas-depleted galaxies as open blue dots. (Right) Residuals to our fit as a function of stellar mass for all non-isolated galaxies (Top, Right) and all isolated galaxies (Bottom, Right). The break in the power-law is shown as a vertical, dashed grey line. The rms scatter of the residuals is listed at the bottom left corner of each panel. We plot a dashed blue line representing the $-1$~dex threshold where we define galaxies to be gas-depleted when located below this threshold. Note that no isolated galaxies are found below a residual of $-1$~dex. We plot non-isolated gas-depleted galaxies as open blue dots. The non-isolated galaxies are slightly more scattered but the primary relation between stellar mass and atomic gas mass is the same for both the non-isolated and isolated galaxies.}
\label{fig_m_hi_m_star}
\end{figure*}

The fraction of baryons in atomic gas, $f_{\rm gas}$, is a strong indicator of the evolutionary state of a galaxy \citep[e.g.,][]{McGaugh:1997fk}. Spheroidal, low mass galaxies in dense environments can be almost completely depleted of HI gas, such as the satellite galaxies of the Local Group \citep[e.g.,][]{Spekkens:2014he}. Low mass galaxies in moderately isolated environments have been shown to form stars slowly and in sparsely distributed regions \citep{vanZee:2000bp, vanZee:2001fm} with star formation becoming more stochastic at low masses \citep{Kauffmann:2014vf}. In fact, the timescale for gas consumption due to star formation alone in gas-rich, irregular, low mass galaxies is $\sim 20$~Gyr \citep{vanZee:2001fm, Roychowdhury:2014jf}. 

Given the timescale for gas depletion due to star formation in low mass galaxies and the fact that isolated low mass galaxies are all forming stars (G12), isolated galaxies should contain a significant amount of baryons in atomic gas. We might also expect gas-depleted low mass galaxies to be located in dense regions and to have undergone environmental processes, as opposed to have experienced purely internal processes that might deplete atomic gas. Therefore, examining the atomic gas fractions of galaxies in isolation compared to denser environments can provide us with insight into how feedback processes and environmental effects regulate the atomic gas supply of galaxies. 

We present the relation between stellar mass and atomic gas fraction in the left panel of Figure \ref{fig_m_star_f_gas} and the relation between stellar mass surface density and atomic gas fraction in the right panel of Figure \ref{fig_m_star_f_gas}.

In Figure \ref{fig_m_star_f_gas} (Left), at stellar masses above $\sim 10^{9.25} M_{\odot}$, galaxies follow a scattered relation with an upper ridgeline on $f_{\rm gas}$ at all stellar masses. At stellar masses below $10^{9.25} M_{\odot}$, the atomic gas fraction appears to saturate at $M_{*} \sim 10^{8.75}$ with $f_{\rm gas} \sim 95$\%. This implies that high mass galaxies are either more efficient at converting their gas into stars or are more effective at preventing gas from cooling than galaxies at low masses. The scale dependency on $f_{\rm gas}$ at low masses is not introduced simply because of our dynamic range in this figure. This fact is made more apparent by Figure \ref{fig_m_hi_m_star}, where a break in the relation between stellar mass and atomic gas mass is clearly observed.

While the median atomic gas fractions for isolated and non-isolated low mass galaxies are similar ($\MedianGasFractionIso$~for isolated galaxies and $\MedianGasFractionSats$~for non-isolated galaxies) isolated galaxies present slightly less scatter in $f_{\rm gas}$ at fixed stellar mass. Indeed, non-isolated galaxies with $M_{*} < 10^{9.25} M_{\odot}$ here and satellite galaxies from the literature \citep[e.g., Galactic dSphs from ][]{Spekkens:2014he}, can inhabit nearly the full range of $f_{\rm gas}$ (from 0.00 to 0.95) while we observe all isolated galaxies with $M_{*} < 10^{9.25} M_{\odot}$ to have $f_{\rm gas} > 0.3$.

We test that the lack of isolated galaxies with $f_{\rm gas} < 0.3$ is not simply due to a sampling effect. Over 10,000 trials, we randomly select the number of isolated low mass galaxies ($N_{\rm isolated} = \NumberLowMassIsoWithGas$) with replacement from all \NumberLowMassAllWithGas~galaxies with $M_* < 10^{\LowMassThreshWithGas} M_{\odot}$. On average, we draw 5 galaxies with $f_{\rm gas} < 0.3$ and we always draw at least 1 galaxy with $f_{\rm gas} < 0.3$. The complete lack of observed isolated galaxies with $f_{\rm gas} < 0.3$ is therefore compelling evidence for the effects of environmental processes on atomic gas regulation.

We observe atomic gas fractions in our low mass isolated sample that vary from 0.30 to 0.99. Therefore isolated galaxies retain a significant fraction of baryons in atomic gas at low masses. {\it The fact that we do not detect isolated low mass galaxies with $f_{\rm gas}$ below 0.3 is a strong constraint on internal feedback processes, since it appears that feedback processes alone in low mass galaxies does not remove all of a galaxy's atomic gas.} We note that the isolated galaxies with upper limit calculations cannot be conclusively ruled out as having lower atomic gas fractions than this limit. Feedback processes are most likely important in driving the large observed scatter in $f_{\rm gas}$ at low masses.

All but one non-isolated galaxy with $M_{*} < 10^9 M_{\odot}$ and $f_{\rm gas} < 0.3$ are observed by the blind ALFALFA survey. For clarity, we have over plotted a dashed grey line in Figure \ref{fig_m_star_f_gas} (Left) which represents the smallest HI mass that overlaps with the NSA catalog and therefore the smallest $f_{\rm gas}$ that the ALFALFA survey could reach. Due to significantly longer exposure times, our observing program can detect galaxies at even lower atomic gas fractions than this grey line. We also over plot a dashed purple line in Figure \ref{fig_m_star_f_gas} (Left) which represents the smallest $f_{\rm gas}$ that our survey could have detected (the minimum HI mass detected by our observations is $10^{\OurMinHIMass} M_{\odot}$). We observe one non-isolated galaxy and two isolated galaxies below the ALFALFA limit. Also see \S\,\ref{subsec_sensitivity} for a discussion of the sensitivity of our detections.

Galaxies with large $f_{\rm gas}$ are either less efficient at converting their gas into stars or these galaxies are accreting enough gas to replenish their star-making fuel supply. As expected from galaxies that follow the Kennicutt-Schmidt relation \citep[see][section 3.2]{Zhang:2009cr} and from previous observations \citep[e.g. ][]{Catinella:2010eo, Cortese:2011hz}, we find that $f_{\rm gas}$ anti-correlates with the surface density distribution of stars. In the right panel of Figure \ref{fig_m_star_f_gas}, we present the relation between stellar mass surface density and atomic gas fraction. We calculate the stellar mass surface density as,

\begin{equation}
\mu_{*}~=\frac{M_{*}}{2 \pi r_{\rm eff}^2},
\end{equation}

\noindent where $r_{\rm eff}$ is the effective radius which encloses 50\% of the optical light. The effect of environment on $f_{\rm gas}$ is essentially removed from this relation, indicating that galaxies may evolve along this relation. The distribution of stellar mass density is clearly a better predictor of $f_{\rm gas}$ than stellar mass alone. The galaxies with the highest atomic gas fractions tend to have the lowest surface density of stars. 

\subsection{Gas Mass Versus Stellar Mass}
\label{subsec_m_star_m_gas}

We present the stellar-to-atomic gas mass relation in the left panel of Figure \ref{fig_m_hi_m_star}. The slope of this relation steepens just below a stellar mass of $10^9 M_{\odot}$. This break in the relation is the cause of the saturation of $f_{\rm gas}$ at low mass in the left panel of Figure \ref{fig_m_star_f_gas}.

We fit a broken power-law to the high and the low mass isolated galaxies only. The fit breaks at $M_{*} =10^{\mhimstarbreak} M_{\odot}$, where we find the two linear fits best intersect and minimize the scatter in the fits. This is a slightly lower stellar mass than our definition of a low mass galaxy. We present our fits both in this panel and in Table \ref{tbl_scaling_fits}. As discussed in \S\, \ref{subsec_sensitivity}, we find no evidence that this break in the power-law is due to a selection effect: We have no evidence that we have missed a significant galaxy population at low stellar mass and high atomic gas masses. This break in the power-law has also been observed by other studies \citep[e.g.,][]{Huang:2012gk}. The break in this relation implies that star formation is either much less efficient in low mass galaxies or the HI gas in low mass galaxies stays hot enough to prevent star formation but not hot enough to ionize the HI.

We inspect the residuals to our fits in the right panel of Figure \ref{fig_m_hi_m_star}. The residual panels demonstrate that scatter towards relatively small atomic gas masses is significant for non-isolated galaxies but not so for isolated galaxies. As discussed in \S\, \ref{subsec_sensitivity}, the non-isolated panel is certainly incomplete at low atomic gas fractions due to ALFALFA survey limits, yet we still observe many galaxies with relatively large negative residuals. In fact, \textit{we only detect non-isolated galaxies at residuals less than $-1$~dex}. All isolated galaxies have either been either been detected in HI or cannot be ruled out at high HI masses via their upper limits. Therefore, we do not detect any isolated galaxies below a residual of $-1$ dex.

Note that $-1$~dex is more than $-3\sigma$ away from our primary relation (see the rms values in each panel of Figure \ref{fig_m_hi_m_star} (Right)) and we plot a dashed blue line in both panels of this figure that represents residuals of $-1$~dex. Below this residual threshold, we define gas-depleted galaxies as open blue dots in this and our remaining figures for reference. We choose $-1$~dex as our residual threshold because we do not observe any galaxies above $1$~dex in either sample and we do not observe any isolated galaxies below $-1$~dex. These panels reinforce the idea that environmental effects preferentially remove atomic gas from non-isolated galaxies. The non-isolated outliers with residuals below $-1$~dex provide evidence that galaxies are depleted of atomic gas due to environmental processes. 

\subsection{Maximum Circular Velocity, Galaxy Mass and Galaxy Size Relations}
\label{subsec_vms_scaling}

\begin{figure*}[t!]
\epsscale{1.2}
\plotone{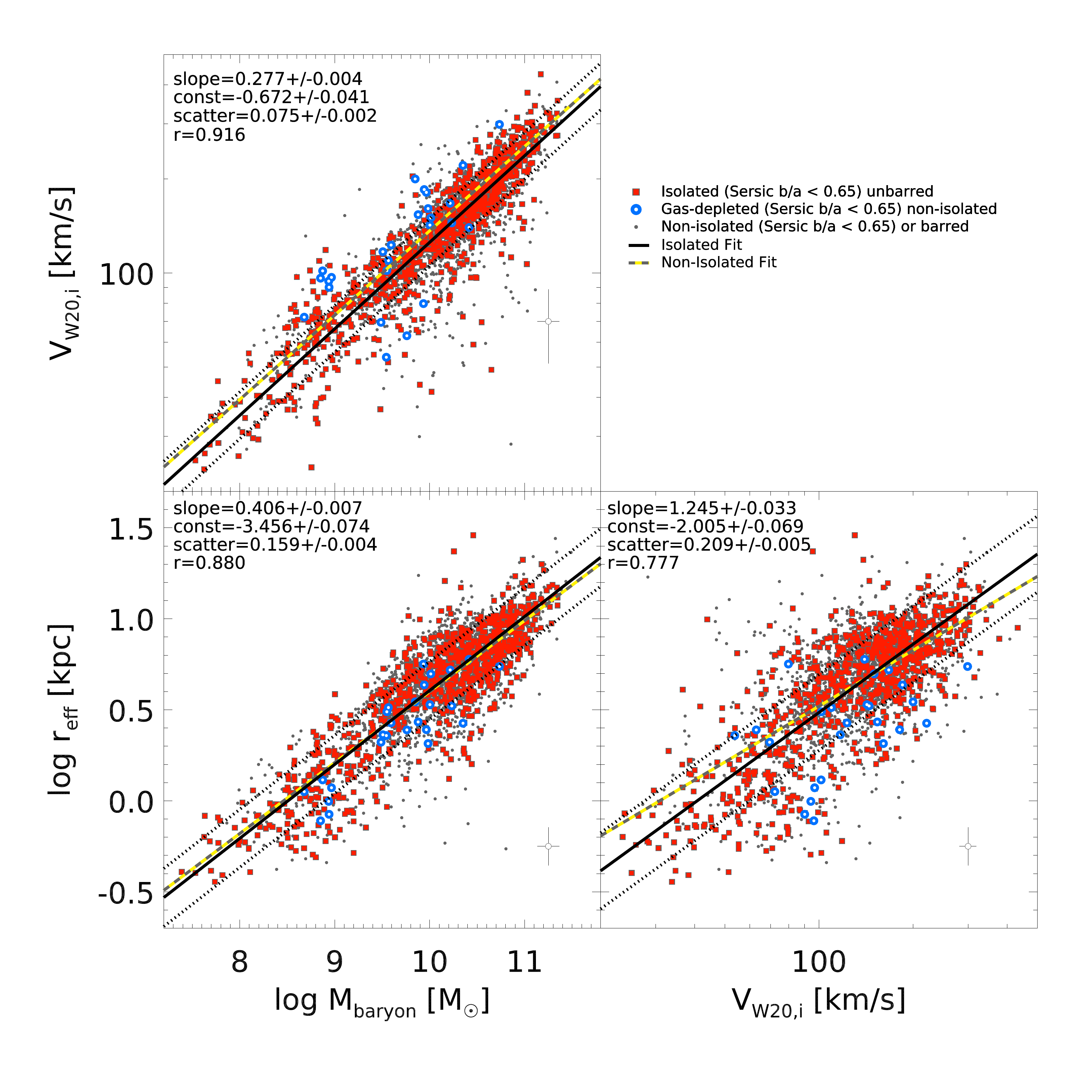}
\caption{Velocity-mass-size scaling relations for $V_{W20,i}$, $M_{\rm baryon}$ and $r_{\rm eff}$. Unbarred, isolated galaxies with axis ratios less than 0.65 are plotted as red squares. Non-isolated galaxies with axis ratios less than 0.65 and barred isolated galaxies are plotted as grey dots for reference. Gas-depleted, non-isolated galaxies with axis ratios less than 0.65 are plotted as open blue dots. The slope, constant, scatter and Pearson correlation rank (r) are listed in the top left corner of each panel. An open data point with representative error bars are plotted in the bottom right corner of each panel. The fit of each relation is plotted as a solid black line with $\pm 1 \sigma$ scatter plotted as dotted black lines. We also plot the relations measured with the non-isolated sample only as dashed yellow lines for reference. The fits are similar for isolated, non-isolated and both samples combined (see Table \ref{tbl_scaling_fits}). The relation between $V_{W20,i}$ and $r_{\rm eff}$ appears to be affected by environment.}
\label{fig_v_r_m}
\end{figure*}

Distinct dark matter halos and their respective disk galaxy properties are related via galaxy scaling relations \citep[e.g.][]{Courteau:2007ek, AvilaReese:2008gz}. These relations are critical for studying models of galaxy formation and evolution \citep[e.g.][]{vandenBosch:2000iw, Dutton:2007he}. We measure the velocity-mass-size scaling relations of our isolated disk galaxies using HI velocity widths, baryonic masses, and optical radii. Here we focus on galaxy scaling relations in isolation and compare our results to galaxies in denser environments.

In each panel of Figure \ref{fig_v_r_m}, we present the velocity-mass-size relations as well as a fit to each relation using only unbarred, isolated galaxies with axis ratios less than 0.65. For comparison, we also plot the fit for non-isolated galaxies only as a yellow dashed line in each figure. As mentioned in \S\,\ref{subsec_quant}, we have removed known barred galaxies from our analysis because their inclination measurements tend to fail. In Table \ref{tbl_scaling_fits}, we present the results of fitting a linear model of $\log y = \alpha \log x + \beta \pm \epsilon$ to our data, where $\epsilon$ represents the intrinsic, random scatter in $y$. We also provide a measurement of the Pearson product-moment correlation coefficient (Pearson r) for each galaxy scaling relation. We provide model fits and Pearson r measurements for 1.) all galaxies, 2.) isolated galaxies and 3.) non-isolated galaxies.

We inspect the velocity-mass-size scaling relations of isolated disk galaxies. For maximum circular velocity measurements, we find that at $V_{W50,i} < 100$\kms the inclination corrected 50\% velocity width can significantly deviate from the galaxy scaling relations of more massive galaxies, therefore we use the inclination corrected 20\% velocity width ($V_{W20,i}$). For mass measurements, $M_{\rm baryon}$ has been shown to minimize the scatter of the BTF for low mass, gas-dominated galaxies \citep[][]{McGaugh:2000hx, McGaugh:2005bc, Stark:2009ks, McGaugh:2012ev}. $M_{\rm baryon}$ should therefore provide the best mass measurement for studying galaxy scaling relations, as opposed to optical luminosity or stellar mass alone. Finally, we study the sizes of our sample using effective radii, $r_{\rm eff}$. Other choices of Petrosian isophotal radii are similarly scattered.

\subsubsection{Velocity versus Baryonic Mass}
\label{subsubsec_mv_relation}
In Figure \ref{fig_v_r_m} (Top, Left) we show the relation between baryonic mass and maximum circular velocity, the BTF. We present two slightly different fits to this relation in Table \ref{tbl_scaling_fits} using mass and velocity as the independent variable in each case. This relation is the most linearly correlated with a Pearson r of \isomvpearsonrank~and only slightly decreases to \nonmvpearsonrank~for the non-isolated sample. The classical Tully-Fisher relation is between optical luminosity and circular velocity while the BTF relates the total baryonic mass of the disk to the gravitational potential of the disk \citep{Tully:1977wu, Mo:2010wea}. When baryonic mass is used instead of luminous mass, galaxies at all masses follow a tight relation at all velocities \citep{McGaugh:2000hx}. This indicates that despite being gas-dominated, low mass, irregular galaxies still follow the same fundamental relation as more massive, stellar mass dominated galaxies \citep{McGaugh:2009by}. Indeed, we observe a very small scatter of \isomvscatter~for the isolated sample and \nonmvscatter~for the non-isolated sample. We do not observe a significant slope change due to environment. Gas-depleted non-isolated galaxies are not preferentially scattered above or below this relation. We plan to examine the BTF relation in more detail in a future paper.

\subsubsection{Radius versus Baryonic Mass}
In Figure \ref{fig_v_r_m} (Bottom, Left) we show the relation between radius and baryonic mass. The relation is linearly correlated with a Pearson r of \isoMrpearsonrank~which decreases to \nonMrpearsonrank~for non-isolated galaxies. This relation is an estimate of the overall density and thus the concentration of a galaxy. \citet{Firmani:2009ho} have shown that disk galaxies move along this relation as they evolve, indicating that we may not observe environmental effects on the baryonic mass to radius relation if radius and baryonic mass are both modified by environmental processes. Indeed, while non-isolated outliers exist, these outliers barely affect the fit, slope and scatter of the relation. We note that gas-depleted non-isolated galaxies are slightly scattered to either smaller radii or larger baryonic masses. Since these galaxies are actually deficient of baryons compared to galaxies of similar stellar mass, we assume that the radii of these gas-depleted galaxies are slightly smaller than the isolated sample.

\subsubsection{Radius versus Velocity}
In Figure \ref{fig_v_r_m} (Bottom, Right) we show the relation between radius and maximum circular velocity. This relation is the least linearly correlated with a Pearson r of \isovrpearsonrank, which decreases to \nonvrpearsonrank~for the non-isolated sample. The radius-velocity relation is related to the specific angular momentum of a galaxy. This relation is the most scattered of the three scaling relations. The scatter may be partially driven by the choice of radius, but relatively large scatter is predicted by models of galaxy formation \citep{AvilaReese:2008gz}. This relation presents the most significant slope change for a scaling relation due to environment, becoming shallower from $\isovrslope \pm \isovrslopeerr$~for isolated galaxies to $\nonvrslope \pm \nonvrslopeerr$~for non-isolated galaxies (also see Table \ref{tbl_scaling_fits} for a list of all fits). The scatter does not change when we fit on the isolated sample compared to the non-isolated sample. The change in slope implies that either the effective radius decreases for non-isolated galaxies at fixed circular velocity, or that circular velocities somehow increase. Non-isolated galaxies may be undergoing environmental effects that are changing the distribution of luminous mass and increasing the half-light radius of our systems. Gas-depleted non-isolated galaxies tend to be slightly more scattered to smaller radii or higher maximum circular velocities. It is not obvious why the gas-depleted non-isolated galaxies would be scattered to higher maximum circular velocities. Since we also observe a similar trend with radius as a function of baryonic mass, we assume that these gas-depleted galaxies tend to have slightly smaller effective radii than the isolated sample.

\subsection{Baryon Fractions}
\label{subsec_baryon_fractions}

\begin{figure*}[t!]
\epsscale{1.19}
\plotone{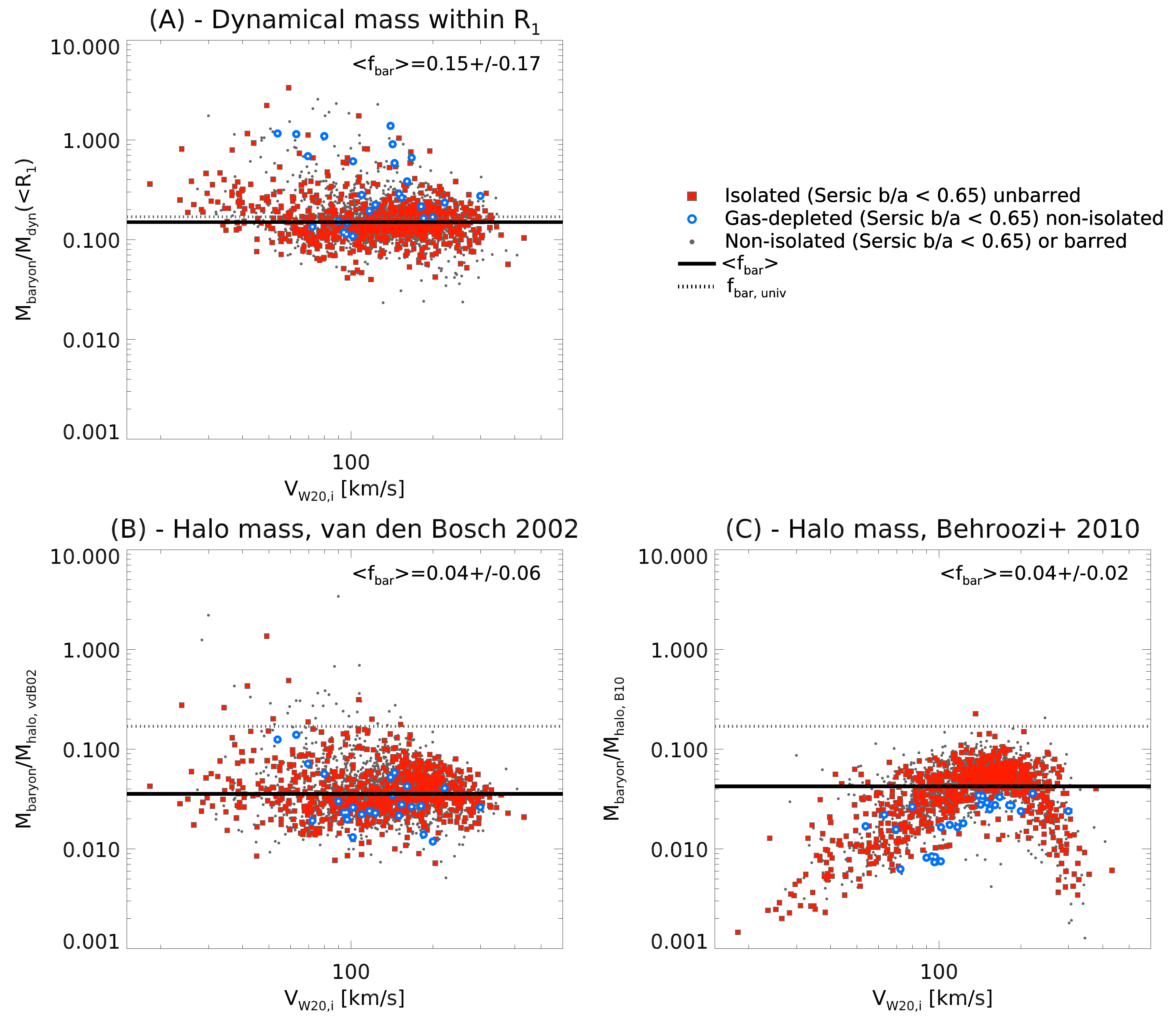}
\caption{Disk (panel A) and halo (panels B and C) baryon fractions versus circular velocity. Unbarred isolated galaxies with axis ratios less than 0.65 are plotted as red squares. Non-isolated galaxies with axis ratios less than 0.65 and barred isolated galaxies are plotted as grey dots for reference. Gas-depleted, non-isolated galaxies with axis ratios less than 0.65 are plotted as open blue dots. We plot the median baryon fraction as solid black horizontal lines. We plot the universal baryon fraction of 0.17 as a dotted horizontal line for reference. Typical uncertainties in $V_{W20,i}$ and $f_{\rm baryon, disk/halo}$, given purely by error propagation, are on the order of $10$~\kms and $0.01$. We also output the median baryon fraction in the top right of each figure. The baryon fractions in panels A and B are calculated using the circular velocities we measure and result in flat relations. This is most likely due to the limitations of using HI circular velocities in low mass galaxies where the HI gas may not probe out to $V_{\rm max}$. The baryon fractions in panel C are measured independently of circular velocity but is the only panel where we find a strong relation with circular velocity. }
\label{fig_baryon_fraction}
\end{figure*}

If the atomic gas supply of galaxies is indeed dependent on environment, we would predict a higher measured baryon fraction in our isolated low mass galaxy sample compared to galaxies that have had their atomic gas depleted by environmental processes. We define the baryon fraction of a galaxy as the total mass of baryons divided by the total mass of the galaxy (baryons plus dark matter). 

The total baryonic mass of a galaxy is relatively straightforward to estimate. However, in calculating a galaxy's baryonic mass, we will always come up short of the actual baryonic mass due to missing contributions from dust, molecular gas and ionized hydrogen. The total galaxy mass (baryons and dark matter) is much more ambiguous. The total galaxy mass can either be estimated as the total dynamical mass directly from the measured galaxy dynamics (hereafter the ``disk baryon fraction") or from indirect estimates of the total halo mass (hereafter the ``halo baryon fraction"). In both cases, we define baryon fraction as:

\begin{equation}
 f_{\rm baryon, disk/halo} = \frac{M_{\rm baryon}}{M_{\rm dyn/halo}}.
 \end{equation}

\noindent In Figure \ref{fig_baryon_fraction}, we present 3 different baryon fractions: 1 disk baryon fraction and 2 halo baryon fractions. For the calculations below, we assume that $V_{W20,i}$ is the maximum circular velocity of our galaxies.

The disk baryon fraction measures the fraction of baryons that have settled into the disk of the galaxy \citep[i.e., the condensed baryon fraction of galaxies, ][]{Zaritsky:2014fr}. For disk baryon fractions, we assume that all of our galaxies are disk galaxies and that $V_{W20,i}$ has been measured at some radius related to the size of the HI disk. We calculate the total dynamical mass, which includes dark matter and baryons and assumes a spherical distribution of mass, as

\begin{equation}
M_{\rm dyn}(<R_{\rm HI}) = \frac{R_{\rm HI}V_{W20,i}^2}{G}~[M_{\odot}].
\label{eq_dyn_mass}
\end{equation}

Since we do not have resolved HI rotation curves, we estimate the HI disk radius. We use the empirically derived relation of \citet{Broeils:1997ut} between $D_{\rm HI}$ and $M_{\rm HI}$ (their Equation 3), where $D_{\rm HI}$ is the HI disk diameter at a surface density of $1 M_{\odot}/ \rm pc^{-2}$,

\begin{equation}
\log{M_{\rm HI}} = 1.96 \log{D_{\rm HI}} + 6.52.
\end{equation}

\noindent This relation is calibrated with spiral and irregular HI rich galaxies with $10^{7.9} < M_{\rm HI} < 10^{10.3}$, (also see \citet{Wang:2014gb, Hallenbeck:2014ku}). We define the HI disk radius as $R_{\rm HI} = D_{\rm HI}/2$.

 \citet{deBlok:2014bm} have shown that 50\% HI velocity widths originate at roughly $5R_{D}$, where $R_{D}$ is the optical disk scale radius. The near equivalence of $5R_{D}$ and $R_{\rm HI}$ has been noted by \citet{deBlok:2014bm}. We compare dynamical masses using these two estimates and find that $R_{\rm HI}$ and $5R_{D}$ produce similar results but $R_{\rm HI}$ results in a less scattered relation between circular velocity and baryon fraction. 
 
We also compare dynamical mass estimates using $R_{\rm HI}$ to dynamical masses using the HI beam radius,

\begin{equation}
r_{\rm beam} = D \rm~sin \it \left(\frac{\theta_{\rm beam}}{\rm 2} \right),
\label{eq_r_beam}
\end{equation}

\noindent where $D$ is the distance to the galaxy and $\theta_{\rm beam}$ is the angular size of the radio beam in 21-cm. The HI beam radius is the maximum radius at which we could have observed HI emission. Since we have not measured the actual radius of the disk's maximum circular velocity, this radius is the maximum radius that we could have possibly measured $V_{W20,i}$. Again, these radii give similar results, but the estimate using $R_{\rm HI}$ has the smallest amount of scatter in the relation below. Therefore we chose to estimate only one disk baryon fraction. Note that if dark matter exists outside of the radius of the 21-cm observations (the beam), or if the rotation velocity profile is increasing beyond this radius and $V_{W20, i} < V_{\rm max} $, then we would estimate a larger baryon fraction. 

We next calculate halo baryon fractions of our isolated galaxies using two different estimates of virialized halo mass. In their models, \citet{vandenBosch:2002kk} (hereafter vdB02) find that the product of the square maximum circular velocity and the disk scale radius provide the best recovery of virial mass when compared to estimates using total galaxy luminosity or maximum circular velocity alone. We therefore calculate virial mass using scale radius and maximum circular velocity via equation 15 of vdB02 as,

\begin{equation}
M_{\rm halo, vdB02} = 2.54 \times 10^{10} M_{\odot} \left( \frac{R_{\rm D}}{\rm kpc} \right) \left( \frac{V_{W20,i}}{100 \rm kms} \right)^2.
\end{equation}

\noindent We assume exponential disks in our galaxies and set $R_{\rm D} = 1.67 r_{\rm eff}$. This is a valid approximation for most of our galaxies where the median S\'ersic index is 1.3.

We also use the stellar mass to halo mass relation of \citet{Behroozi:2010ja} (hereafter B10). This relation between stellar mass and halo mass has been calibrated by taking advantage of the abundance matching technique. This technique assumes that stellar mass is directly related to halo mass with scatter less than 0.2 dex at all masses. For this halo mass estimate, B10 assume that the stellar mass function accurately maps onto halo mass function, even at the lowest masses where galaxies' baryonic mass are often dominated by atomic gas \citep{Papastergis:2012cb} and where many galaxies may be missed due to the surface brightness limitations of the input stellar mass function \citep{Blanton:2005gz}. B10 abundance match SDSS stellar mass functions to halo mass functions with virial masses at 337 times the mean background density at z = 0 (the same virial mass definition as vdB02). We use B10 Equation 21 with their free $\mu$ and $\kappa$ model for $0<z<1$, we call this mass estimate $M_{\rm halo, B10}$.

In Figure \ref{fig_baryon_fraction}, we present each of the three different estimates of baryon fractions. Uncertainties on $V_{W20,i}$ and $f_{\rm baryon}$ are typically on the order of 10\,\kms and 0.01 respectively, given \textit{purely} by error propagation. We list the median baryon fractions with rms scatter at the top right of each panel and as a yellow dashed horizontal line.

The first panel (A) of Figure \ref{fig_baryon_fraction} presents a disk baryon fraction as a function of circular velocity. The next two panels (B and C) present halo baryon fractions as a function of circular velocity and stellar mass. For reference, we plot the universal baryon fraction $f_{\rm baryon} = 0.17$ in each panel as a horizontal dotted line. We discuss each panel of Figure \ref{fig_baryon_fraction}. We visually inspect the SDSS images of outliers in each of the three panels. 

\begin{enumerate}[(A)]
\item We measure a median disk baryon fraction of $\FBARYONDISKFRACTION$. This measurement is similar to the universal baryon fraction. We assume this is because we are measuring the dynamical mass at $R_{\rm HI}$ rather than the virial mass at $R_{\rm vir}$, so we are not measuring a significant mass contribution from the dark matter halo. Galaxies at large baryon fractions are largely driven by mis-measured dynamical masses due to bars that have gone un-identified in Galaxy Zoo. These bars dramatically affects our baryon fractions due to the way we estimate dynamical masses using $R_{\rm HI}$ and $V_{W20,i}$. Galaxies at very low baryon fractions are mostly LSB galaxies. It also appears that the relation between HI mass and $R_{\rm HI}$ begins to break down at small circular velocities or that low mass galaxies have larger baryon fractions in their galactic disks. Gas-depleted galaxies tend to lie at large baryon fractions, indicating that the dynamical masses can be severely underestimated when a galaxy is gas-depleted. 

\item The halo baryon fraction estimate using the virialized halo mass of \citet{vandenBosch:2002kk} who fit the relation between $R_{d}V^2_{\rm max}$ and $M_{\rm vir}$. The median value is $\FBARYONHALOFRACTIONVDB$. Outliers at large baryon fractions are due to barred galaxies not identified by the Galaxy Zoo catalog, possible mergers, and irregular morphologies. Galaxies at very low baryon fractions are mostly LSB galaxies. Galaxies at small circular velocities do not show the upturn in baryon fraction that appears in panel A, they remain constant at nearly all circular velocities with significant scatter and a slight decrease in baryon fraction at high masses. Gas-depleted galaxies appear unperturbed from the primary relation.

\item The halo baryon fraction estimate using the stellar mass to halo mass abundance matching of \citet{Behroozi:2010ja} recovers the only obvious relation between circular velocity and baryon fraction. While the dynamical masses estimated in panel A and the halo masses estimated in panel B are calculated using a power law relation with maximum circular velocity, the B10 stellar mass to halo mass relation results in somewhat of an exponential relation with maximum circular velocity (not shown here). This relation flattens out at small circular velocities (below $\sim 70$~\kms and asymptotes at large circular velocities (above $200$~\kms). For galaxies with $80 < V_{W20,i} <100$~\kms, the vdB02 and B10 baryon fractions are extremely similar because this is where the relations between maximum circular velocity and halo mass overlap. The B10 measurements of baryon fractions imply that both low mass and high mass galaxies are severely deficient of baryons compared to both the cosmic value. The median baryon fraction using this halo mass estimate is $\FBARYONHALOFRACTIONB$, which is equal to, but less scattered than the estimate in panel B. Gas depletion in galaxies does not affect this estimate of halo mass, so gas-depleted galaxies simply lie at smaller baryon fractions than the rest of the sample.

\end{enumerate}

In the cases above that estimate baryon fractions using halo mass, the measured baryon fraction is depleted compared to the universal baryon fraction. As we mentioned earlier, the baryon fractions of our galaxy sample must be somewhat depleted since we have not detected all of the ionized gas, molecular gas and dust in these galaxies. We assume that our atomic gas observations allow us to make reliable estimates of total baryonic mass. We discuss the impact of molecular gas on our results further in \S\ \ref{subsec_gfhs}.

The scaling relation between maximum circular velocity and baryonic mass we measure in \S\, \ref{subsec_vms_scaling} suggests a weak correlation between maximum circular velocity and baryon fraction: We measure $M_{\rm baryon} \propto V^{\isovmslope}$ and $r \propto V^{1/ \isovrslope}$ in \S\, \ref{subsec_vms_scaling}. If we assume that $M_{\rm dynamic} \propto rV^{2}$ then $M_{\rm baryon}/M_{\rm dyn}\propto V^{\isofbaryonvpower}$. We do not observe such a slope in the three panels above, i.e. this derived slope implies a smaller baryon fraction at low masses and a larger baryon fraction at high masses, but we observe a mostly constant baryon fraction in panels A and B above.

Since the precise value of $r$ in our dynamical mass estimate is uncertain, we could instead assume a direct relation between total virialized mass and maximum circular velocity. We compare to the results of N-body simulations from \citet{Klypin:2011bd} who find that $M_{\rm vir} \propto V_{\rm max}^{3.16}$ for distinct halos (see their equation 8). Given the previous paragraph, this would imply a roughly constant baryon fraction at all velocities, as in cases A and B above.

In general, we find that baryon fraction is independent of environment. The lack of environmental dependence on baryon fraction is not surprising given our results in \S\,\ref{subsubsec_mv_relation}. We note that the scatter in baryon fraction doubles when the non-isolated galaxy sample is included. We find no evidence for severe baryon depletion in low mass galaxies when we attempt to directly measure dynamical or halo mass, as opposed to the baryon deficiency using the abundance matching result. This result motivates further research into the most accurate measurement of halo mass in the lowest galaxy mass regime.

\section{Discussion and Comparison to Similar Studies}
\label{sec_discussion}

We discuss our results in context with similar studies, specifically in terms of environmental effects. We focus on results from the ALFALFA survey and other work that has inspected comparable HI observations of late-type galaxies at low redshift over similar galaxy mass ranges.

\subsection{Gas Fractions and the Gas to Stellar Mass Relation}
\label{subsec_gfhs}

Our relation between $M_{*}$ and $M_{HI}$ agrees well with \citet{Zhang:2009cr}, who use the homogeneous HyperLeda catalog and the SDSS DR4 to inspect the difference in HI masses for 800 metal-rich and metal poor galaxies as a function of fixed stellar mass between $10^{7.5} M_{\odot}$ and $10^{11} M_{\odot}$. These authors find an offset between metal-poor and metal-rich galaxies with metal-poor galaxies showing higher HI gas fractions than their metal-rich sample. These results suggest that galaxies with higher HI gas fractions are less evolved (at least chemically), indicating that our isolated low mass sample at high atomic gas fractions are the least evolved galaxies at this mass scale. It may also be that these gas-rich galaxies have simply accreted a significant amount of low metallicity gas.

The atomic gas content of SDSS DR 7 galaxies have been studied extensively using the \ALFALFA~dataset in \citet{Huang:2012gk} and \cite{Huang:2012bv}, hereafter H12a and H12b respectively, as well as in \citet{Maddox:2014fp}. At low masses, H12a note a similar break in the relation of $f_{\rm gas}$ to $M_{*}$ at $M_{*} = 10^9 M_{\odot}$. They suggest this break is consistent with the effects of AGN feedback on the atomic gas content in massive galaxies. Recently, evidence of AGN has been found in galaxies with stellar mass as low as $10^{8.16} M_{\odot}$ \citep{Reines:2013bp}, so the effect of AGN feedback must be less pronounced in low mass galaxies for this to be true.

Why such a pronounced break in the stellar mass to atomic gas mass relation exists at this stellar mass is not entirely clear and is not predicted by existing models of galaxy formation \citep{Maddox:2014fp}. Since the break in the relation is independent of environment, it is most likely due to internal processes of galaxy formation and star formation. Given the fact that we detect small fractions of baryons in these galaxies, we could assume that some significant component of baryons have gone undetected.

We estimate the amount of missing gas from high mass galaxies using the low mass end of the relation between stellar mass and atomic gas mass measured in \S\, \ref{subsec_m_star_m_gas}. We find that high mass galaxies would need roughly four times the measured HI mass in order to account for the break in the power law. While it may be that high mass, late-type galaxies contain a significant amount of molecular hydrogen (e.g. the Milky Way has roughly a three-to-one $M_{HI}$ to $M_{H_2}$ ratio \citep{Draine:2011tr} and M94 has a one-to-one $M_{HI}$ to $M_{H_2}$ ratio \citep{Leroy:2009di}), the molecular gas mass in high mass spiral galaxies appears at most to be comparable to the neutral atomic gas mass \citep{McGaugh:1997fk, Leroy:2008jk, Saintonge:2011hz}. The amount of molecular hydrogen required to makeup this mass deficit would be extreme \citep[e.g.,][]{Leroy:2008jk} but combining this molecular gas with ionized gas may significantly reduce the discrepancy. 

While the ionized gas content of galaxies is not well-known, the molecular gas content of galaxies has been studied extensively \citep[e.g.,][]{Kenney:1989dg,Young:1991hh,McGaugh:1997fk, Welch:2010in}. For instance, the AMIGA project has studied the $M_{HI}$ to $M_{H_2}$ ratio of isolated galaxies \citep[see Fig. 15 and 16 of ][]{Lisenfeld:2011es}. Their results imply that galaxies must be strongly interacting in order to increase the ratio of molecular gas mass to neutral gas. This is most likely because the HI gas disk is less tightly bound than the molecular gas and is stripped first by environmental processes. We would not expect a significant number of such strongly interacting galaxies in our non-isolated sample due to the restriction that the HI beam is not confused between sources.

We estimate the molecular gas mass of our galaxies using H$\alpha$ flux in the SDSS fiber to estimate the star formation rates (SFR) for our galaxies. Since the SDSS fiber aperture is smaller than each galaxy, we correct for the missing H$\alpha$ flux by computing the ratio of the r-band flux in a fiber aperture to the total r-band flux of the galaxy. To estimate molecular gas mass, we assume a relation between SFR and $M_{H_2}$ (see Equation 1 of \citet{McGaugh:2015eu}). It may be that star formation rates are affected by environment, so this exercise may reveal something about environmental effects on our galaxy sample \citep[e.g.,][]{Perea:1997vp}. 

We find the $M_{H_2}$ to $M_{HI}$ ratio of our sample lies between 0.04 and 0.08 and is highly scattered but fairly constant at all masses. We re-run our analysis and fitting algorithms and find that this estimate of molecular gas mass does not impact our results significantly. As expected, we have fewer gas depleted non-isolated galaxies and the slopes of our relations change slightly. While the star formation efficiency of galaxies may not be significantly effected by environment \citep{Casasola:2004iq}, the molecular gas fraction probably is. Therefore an empirical estimate of molecular gas based on star formation efficiency arguments won't pickup on environmental effects if this was true. Also, we focus our work on low mass galaxies and the molecular gas fractions of low mass galaxies are observed to be much lower given their low surface brightness and lower HI surface densities \citep{McGaugh:2012ev, Schruba:2012bb}. We plan to study the effect of environment on SFRs in a future work.

At high masses our data agree well with the GASS survey \citep{Catinella:2010eo}, with their non-detections tracing much lower gas fractions than our sample. The GASS observing strategy is to observe until an atomic gas fraction threshold is reached. Their work can offer insight into non-isolated galaxy evolution at higher gas masses. We are focused on the low mass regime so we forgo a detailed comparison to the GASS survey sample, but we note that GASS high mass non-isolated galaxies are gas-depleted relative to our gas-rich sample.
\subsection{The Gas Content of Galaxies in Low Density Environments}

H12a has shown that galaxies with $M_{*} > 10^9 M_{\odot}$ in the denser Virgo Cluster environment \citep{Cortese:2011hz} are gas-depleted compared to the primary, gas-rich ALFALFA sample (see their figure 2c). \citet{Cortese:2011hz} claim that by comparing different models of Virgo dwarfs, they can determine that ram-pressure stripping is the cause of gas depletion in Virgo spirals as opposed to slow starvation. What is interesting about our results is that despite the \ALFALFA~sample being so gas-rich, we identify a small population of non-isolated galaxies that are noticeably gas-depleted compared to isolated galaxies. It would be interesting to compile a deep HI survey of non-isolated galaxies from the SDSS and probe this relation down to comparable atomic gas fractions as in \citep{Cortese:2011hz}.

While some effort has been made to study the HI content of void galaxies \citep[e.g.,][]{Kreckel:2012it}, our definition of isolated means not all isolated galaxies are located in voids and not all void galaxies are isolated. Therefore results from studies of void galaxies may not be as useful as a direct comparison to our isolated galaxies. Regardless, \citet{Kreckel:2012it} find that low mass void galaxies with stellar mass between $10^7 M_{\odot}$ and $10^9 M_{\odot}$ are \textit{not} preferentially gas-rich. Our results suggest that depending on their definition, void galaxies may be still be affected by their local environment, especially at such low masses. 

\subsection{Scaling Relations}

Our velocity-mass-size scaling relations agree well with similar studies of gas-rich spirals at higher masses \citep{Courteau:2007ek, AvilaReese:2008gz, Tollerud:2010bq, Hall:2012hh}. Since we find that environment can affect the atomic gas fractions of galaxies, it would be logical that the scaling relations of these galaxies are also affected. While galaxy scaling relations become more scattered when non-isolated galaxies are included, we find that the galaxy scaling relations' slopes are not significantly perturbed compared to our isolated gas-rich sample. 

It may not be that galaxies move perpendicular to scaling relations as they undergo environmental processes. \citet{Dutton:2010ci} suggest that observed disk galaxies may evolve parallel to their scaling relations, which may be the case for galaxy evolution due to environmental processes. 

For example, as a galaxy is stripped of atomic gas, it may not form stars at large radii where the binding energy is the lowest and the atomic gas is the most depleted. With its fuel reservoir depleted, a galaxy may also form fewer stars overall than a galaxy that has not been stripped of atomic gas. Therefore, as a galaxy decreases in radius due to environmental processes it may also decrease in stellar mass, moving parallel to the scaling relation between effective radius and baryonic mass. In summary, the effect of environment on galaxies may be difficult to measure using the velocity-mass-size scaling relations.

It may also be that we do not observe a significant environmental dependence in galaxy scaling relations because the ALFALFA survey does not measure the scaling relations of gas-deficient disk galaxies due to the nature of a flux limited survey. In the densest environments, such as in rich group environments \citep{Catinella:2013ej} and dense cluster environments \citep{Kenney:1989dg}, the HI deficiency of galaxies increases dramatically as a function of radius from the X-ray center of the cluster \citep{Boselli:2006ei}. \citet{FernandezLorenzo:2013ez} find evidence that massive isolated spiral galaxies with $10^10 < M_* < 10^11 M_{\odot}$ are larger than galaxies in denser environments at fixed mass.

When S0 galaxies are compared to the Tully-Fisher relation of spirals, an offset and an increase in scatter are observed \citep{Neistein:1999io, Hinz:2003bb, Bedregal:2006jp, Davis:2011bg, Williams:2010hn}. It has also been suggested that when galaxies are either disturbed, compact or merging they do not follow the Tully-Fisher relation and the circular velocity must be added to the non-circular motions of the HI gas \citep{Weiner:2006cz, Kassin:2007ey, Kassin:2014jp}. Taking into account the velocity dispersion and concentration of a galaxy, disk and spheroidal galaxies follow similar scaling relations \citep{Catinella:2012ea}. If spheroidal galaxies are transformed by their environment, then galaxies in groups and clusters should systematically vary from the scaling relations presented in this work. Indeed, \citet{Cortes:2008cy} claim that environmental effects have decreased HI line widths in Virgo cluster galaxies. However, the most noticeable environmental effect on the galaxy scaling relations is that gas-depleted non-isolated galaxies tend to be scattered to smaller effective radii than isolated galaxies - indicating again that we are not sensitive to environmental effects on circular velocity or that $V_{W20,i}$ is also encapsulating noncircular motion.
\subsection{Baryon Fractions}

We find that baryon fraction is independent of circular velocity in all cases except $M_{\rm halo, B10}$. The baryon fraction calculated using $M_{\rm halo, B10}$ shows a significant dependence on circular velocity because this estimate is based on stellar-to-halo mass abundance matching. This result is similar to \citet{Papastergis:2012cb}, who perform abundance matching with total baryonic masses using \ALFALFA~HI masses. These authors find that the depletion of baryons at low halo masses is only mildly alleviated when total baryon mass is used in the abundance matching technique. It has been suggested that LSB galaxies may significantly change abundance matching results by changing the stellar mass function that goes into matching the halo mass function \citep{Ferrero:2012bt}.

Our other baryon fraction estimates are at odds with the abundance matching technique which suggest baryon fractions are dependent on circular velocity and are much more significantly depleted of baryons. The flat relations we measure between circular velocity and baryon fractions at low circular velocities suggests that our HI velocity width measurement is not probing deep into the dark matter halo of low mass galaxies \citep{Ferrero:2012bt}. This could be confirmed by resolved HI synthesis maps. If the HI rotation curves do not reach the maximum circular velocity of the dark matter halo \citep[e.g., several examples are available from ][]{Swaters:2009ku}, our $V_{W20,i}$ measurement would underestimate halo mass. This effect would lead to a flattening of the relation.

\citet{Blanton:2008il} find a weak relation between baryon fraction as a function of luminosity at low masses. \citet{Zaritsky:2014fr} estimate the disk baryon fraction for a similar sample of gas-rich galaxies using the mass estimate of \citet{Bullock:2001hp} and find a baryon fraction of 0.07. \citet{McGaugh:2009by} calculate the fraction of detected baryons to range from 0.83 for clusters to 0.0003 for the lowest mass dwarf galaxies. Our results do not indicate such a significant depletion of baryons in isolated galaxies with stellar mass $10^7 < M_{*} < 10^9 M_{\odot}$, nor do we note a significant trend towards low baryon fractions except in the B10 abundance matching estimate.

\section{Summary}
\label{sec_summary}

We present new measurements of HI gas masses for \NNewIso~isolated galaxies with stellar masses between $10^7$ and $10^{9.5} M_{\odot}$. We apply the same isolation criterion to our low mass sample as G12 and extend this isolation criteria to galaxies at higher masses. We re-measure the HI masses and velocity widths of galaxies observed with the ALFALFA survey in order to create a large, homogeneous data set. Our HI data set represents the largest of its kind with such a rigorously defined isolation criteria.

Using these data, we study the baryon content of galaxies in isolated environments, specifically at low masses. We inspect the relation between galaxy stellar mass and atomic gas mass in isolated environments versus the relation in non-isolated galaxies (i.e. denser environments). We also examine the relations between velocity, mass and size. Finally, we determine the disk baryon fractions and two estimates of halo baryon fractions for our galaxies. Our primary results are as follows:

\begin{itemize}
\item{We either place all isolated low mass galaxies with stellar masses between $10^7$ and $10^{9.5} M_{\odot}$ at high atomic gas fractions or we cannot rule out high atomic gas fractions via upper limits. Isolated galaxies have a median atomic gas fraction $f_{\rm gas} = \MedianGasFractionIso$ and atomic gas fractions always greater than 0.3.}
\item{We use a largely flux-limited sample of non-isolated low mass galaxies to derive a similar median atomic gas fraction as isolated galaxies. We measure non-isolated atomic gas fractions as low as 0.15, which is limited by the HI-rich galaxies of the ALFALFA survey.}
\item{We observe a break in the relation between stellar mass and atomic gas mass at $10^{\mhimstarbreak} M_{\odot}$. Both isolated and non-isolated galaxies follow this relation.}
\item{Only non-isolated galaxies are depleted in atomic gas by more than $-1$~dex compared to the atomic gas mass predicted by the stellar mass to atomic gas mass relation.}
\item{Environment does not significantly affect the linear correlations between velocity, mass and size. Gas-depleted non-isolated galaxies tend to be scattered to smaller effective radii.}
\item{Isolated galaxies follow a steeper relation between velocity and radius than non-isolated galaxies.}
\item{The baryon fraction of isolated galaxies is independent of environment: we measure a median baryon to dynamical mass fraction of $f_{\rm baryon, disk} = \FBARYONDISKFRACTION$ and median baryon to halo mass fractions of $f_{\rm baryon, halo} = \FBARYONHALOFRACTIONVDB$ using the results of semi-analytic models and $f_{\rm baryon, halo} = \FBARYONHALOFRACTIONB$ using abundance matching.}
\end{itemize}

The above results improve our understanding of the baryon content of low mass galaxies and offer us some insight into the effect of environment in this stellar mass regime. Extending our analysis to galaxies below our stellar mass limit of $10^7 M_{\odot}$ requires deeper spectroscopic surveys (e.g., DESI) and deeper all-sky HI surveys (e.g., SKA). The flat relation we measure between circular velocity and baryon fraction motivates resolved, follow-up HI synthesis maps of our low mass galaxy sample in order to determine the true maximum circular velocity of our systems and the radius where we measure this velocity. We plan to study the spatial distribution and the kinematics of baryons in our sample using integral field unit observations.
\begin{acknowledgements}

It is our pleasure to thank Ana Bonaca, Duncan Campbell, Susan Kassin, Allison Merritt, Erik Tollerud, Frank van den Bosch, Benjamin Weiner and Andrew Wetzel for helpful discussions and comments on this manuscript. We would like to acknowledge the referee for their time and constructive comments on this manuscript. We also thank Martha Haynes for allowing us to view the most current unreleased ALFALFA dataset as well as the entire ALFALFA team for their hard work and for providing reduced HI spectra and catalogs for the community. JB acknowledges support from the Gruber Foundation and the National Science Foundation Graduate Research Fellowship Program. MB was supported by NSF-AST-1109432 and NSF-AST-1211644.

The Arecibo Observatory is operated by SRI International under a cooperative agreement with the National Science Foundation (AST-1100968), and in alliance with Ana G. M\'{e}ndez-Universidad Metropolitana, and the Universities Space Research Association. The National Radio Astronomy Observatory is a facility of the National Science Foundation operated under cooperative agreement by Associated Universities, Inc. Funding for SDSS-III has been provided by the Alfred P. Sloan Foundation, the Participating Institutions, the National Science Foundation, and the U.S. Department of Energy Office of Science. The SDSS-III web site is http://www.sdss3.org/. This research has made use of NASA's Astrophysics Data System and Matplotlib \citep{Hunter:2007ih}. This material is based upon work supported by the National Science Foundation Graduate Research Fellowship Program under Grant No. DGE-1122492. Any opinions, findings, and conclusions or recommendations expressed in this material are those of the author(s) and do not necessarily reflect the views of the National Science Foundation.

\end{acknowledgements}



\newpage
\clearpage

\begin{deluxetable}{cccccccccc}
\tabletypesize{\tiny}
\tablecaption{Sample of Our High Quality HI Measurements and Environments for Isolated Low Mass Galaxies}
\tablewidth{0pt}
\tablehead{
\colhead{NSA ID} &
\colhead{$\alpha$ (J2000.0)} &
\colhead{$\delta$ (J2000.0)} &
\colhead{Distance} &
\colhead{$M_{r}$} &
\colhead{$W_{50}$} &
\colhead{$S_{21}$} &
\colhead{$M_{\rm HI}$} &
\colhead{$M_{*}$} &
\colhead{$d_{\rm host}$}\\
\colhead{  } &
\colhead{[h$\,$:$\,$m$\,$:$\,$s]} &
\colhead{[$^\circ\,$:$\,'\,$:$\,''$]} &
\colhead{[Mpc]} &
\colhead{[mag]} &
\colhead{[km s$^{-1}$]} &
\colhead{[Jy km s$^{-1}$]} &
\colhead{[$10^9 M_{\odot}$]} &
\colhead{[$10^9 M_{\odot}$]} &
\colhead{[Mpc]}\\
\colhead{(1)} &
\colhead{(2)} &
\colhead{(3)} &
\colhead{(4)} &
\colhead{(5)} &
\colhead{(6)} &
\colhead{(7)} &
\colhead{(8)} &
\colhead{(9)} &
\colhead{(10)}}
\startdata
162 & 09:58:30.23 & +00:02:42.72 &  30.23$\pm  1.89$ & -13.34 &           50$\pm           5$ &         0.49$\pm        0.03$ &        0.105$\pm       0.029$ &        0.016$\pm       0.007$ &  2.15 \\
1443 & 12:23:09.98 & +00:25:37.76 &  35.36$\pm  2.00$ & -13.61 &           32$\pm          28$ &         0.14$\pm        0.03$ &        0.041$\pm       0.019$ &        0.043$\pm       0.020$ &  1.69 \\
5109 & 22:30:36.80 & -00:06:37.01 &  21.26$\pm  1.86$ & -13.40 &           55$\pm           3$ &         0.63$\pm        0.02$ &        0.067$\pm       0.024$ &        0.008$\pm       0.004$ &  6.28 \\
5378 & 23:06:39.16 & +00:34:41.02 &  68.13$\pm  2.39$ & -15.86 &          104$\pm           7$ &         0.95$\pm        0.03$ &        1.039$\pm       0.166$ &        0.273$\pm       0.126$ &  1.77 \\
5424 & 23:14:19.43 & +00:10:59.36 &  60.30$\pm  1.75$ & -15.67 &          108$\pm           7$ &         0.47$\pm        0.02$ &        0.405$\pm       0.058$ &        0.612$\pm       0.282$ &  2.43 \\
5449 & 23:28:12.31 & -01:03:46.37 &  35.30$\pm  2.14$ & -13.90 &           64$\pm           2$ &         0.78$\pm        0.02$ &        0.229$\pm       0.059$ &        0.020$\pm       0.009$ &  1.51 \\
5720 & 00:13:44.02 & +00:22:18.46 &  56.50$\pm  1.75$ & -14.63 &           29$\pm           2$ &         0.45$\pm        0.02$ &        0.337$\pm       0.049$ &        0.097$\pm       0.045$ &  1.87 \\
6186 & 00:57:56.60 & +00:52:08.56 &  30.17$\pm  2.86$ & -13.91 &           58$\pm           2$ &         0.70$\pm        0.02$ &        0.150$\pm       0.059$ &        0.051$\pm       0.024$ &  3.26 \\
6368 & 01:09:07.85 & +01:07:15.89 &  15.97$\pm  1.64$ & -13.03 &           40$\pm           1$ &         1.02$\pm        0.02$ &        0.061$\pm       0.026$ &        0.015$\pm       0.007$ &  1.69 \\
6412 & 01:13:45.98 & +00:56:44.13 & 104.68$\pm  2.14$ & -16.92 &          131$\pm          21$ &         0.30$\pm        0.02$ &        0.766$\pm       0.135$ &        0.788$\pm       0.363$ &  3.63 \\
6748 & 01:26:05.01 & +00:19:00.68 &  24.85$\pm  2.00$ & -14.26 &          117$\pm           1$ &         3.17$\pm        0.03$ &        0.461$\pm       0.152$ &        0.088$\pm       0.040$ &  1.63 \\
6949 & 01:55:59.76 & -00:11:08.02 &  51.69$\pm  1.75$ & -15.74 &           90$\pm          16$ &         0.33$\pm        0.02$ &        0.208$\pm       0.039$ &        0.323$\pm       0.149$ &  3.48 \\
7167 & 02:18:08.02 & +00:45:29.39 &  39.84$\pm  2.32$ & -14.85 &          116$\pm           1$ &         0.99$\pm        0.02$ &        0.371$\pm       0.091$ &        0.139$\pm       0.064$ &  3.57 \\
8331 & 00:49:22.10 & +15:31:33.00 &  76.64$\pm  2.89$ & -15.81 &          126$\pm           4$ &         0.77$\pm        0.02$ &        1.059$\pm       0.178$ &        0.400$\pm       0.184$ &  2.18 \\
8333 & 00:56:42.19 & +14:12:45.12 & 156.79$\pm  2.14$ & -17.68 &           88$\pm          37$ &         0.67$\pm        0.10$ &        3.883$\pm       1.242$ &        2.932$\pm       1.350$ &  3.21 \\
...\\
\enddata
\tablecomments{Properites of 15 isolated, low mass NSA galaxies that have been observed in HI. Columns are (1) the NASA Sloan Atlas v0.1.2 (NSA) identifier for all galaxies which is unique to this iteration of the NSA catalog, (2) right ascension, (3) declination, (4) heliocentric distance, (5) absolute r-magnitude, (6) width of the 21-cm line at 50 per cent flux, (7) integrated flux of the 21-cm line, (8) HI mass, (9) stellar mass and (10) distance to the nearest massive host. This table is published in its entirety in the electronic edition of the Astronomical Journal, A portion is shown here for guidance regarding its form and content.}
\label{tbl_hienv}
\end{deluxetable}

\begin{deluxetable}{cccccc}
\tabletypesize{\tiny}
\tablecaption{HI Observation Details}
\tablewidth{0pt}
\tablehead{
\colhead{Run ID} &
\colhead{Avg. Integration Time [s]} &
\colhead{N Reduced} &
\colhead{N High Quality} &
\colhead{N Non-Detections} &
\colhead{N Bad Baselines}}
\startdata
GBT2008 & 802 & 184 & 141 & 43 & 12\\
AO2005 & 796 & 32 & 32 & 0 & 3\\
AO2013 & 459 & 83 & 80 & 3 & 8\\
AO2014a & 499 & 51 & 47 & 4 & 0\\
AO2014b & 571 & 21 & 17 & 4 & 0
\enddata
\tablecomments{Observing run identifier, average integration time, number of reduced spectra, number of high quality spectra, number of non-detections, number of bad baselines for all of our HI observations. Note that GBT has a smaller dish diameter than Arecibo, therefore integration times are not directly comparable.}
\label{tbl_obs}
\end{deluxetable}

\begin{deluxetable}{cccccccc}
\tabletypesize{\tiny}
\tablecaption{Scaling Relations with $\log y = \alpha \log x + \beta \pm \epsilon$}
\tablewidth{0pt}
\tablehead{
\colhead{sample} &
\colhead{x} &
\colhead{y} &
\colhead{$\alpha$ (Slope)} &
\colhead{$\beta$ (Constant)} &
\colhead{$\epsilon$ (Scatter in $\log y$)} &
\colhead{Pearson r} &
\colhead{Figure}}
\startdata
isolated only & $M_{*} < 10^{\mhimstarbreak} M_{\odot}$ & $M_{\rm gas}$ & \mhimstarlowslope $\pm$ \mhimstarlowslopeerr & \mhimstarlowconst $\pm$ \mhimstarlowconsterr & \mhimstarlowscatter $\pm$ \mhimstarlowscattererr & N/A & \ref{fig_m_hi_m_star} \\
isolated only & $M_{*} \ge 10^{\mhimstarbreak} M_{\odot}$ & $M_{\rm gas}$ & \mhimstarhighslope $\pm$ \mhimstarhighslopeerr & \mhimstarhighconst $\pm$ \mhimstarhighconsterr & \mhimstarhighscatter $\pm$ \mhimstarhighscattererr & N/A &\ref{fig_m_hi_m_star} \\
isolated only & $M_{\rm baryon}$ & $r_{\rm eff}$ & \isoMrslope $\pm$ \isoMrslopeerr & \isoMrconst $\pm$ \isoMrconsterr & \isoMrscatter $\pm$ \isoMrscattererr & \isoMrpearsonrank & \ref{fig_v_r_m} \\
non-isolated only & $M_{\rm baryon}$ & $r_{\rm eff}$ & \nonMrslope $\pm$ \nonMrslopeerr & \nonMrconst $\pm$ \nonMrconsterr & \nonMrscatter $\pm$ \nonMrscattererr & \nonMrpearsonrank & N/A \\
all & $M_{\rm baryon}$ & $r_{\rm eff}$ & \allMrslope $\pm$ \allMrslopeerr & \allMrconst $\pm$ \allMrconsterr & \allMrscatter $\pm$ \allMrscattererr & \allMrpearsonrank & N/A \\
isolated only & $V_{W20,i}$ & $r_{\rm eff}$ & \isovrslope $\pm$ \isovrslopeerr & \isovrconst $\pm$ \isovrconsterr & \isovrscatter $\pm$ \isovrscattererr & \isovrpearsonrank & \ref{fig_v_r_m} \\
non-isolated only & $V_{W20,i}$ & $r_{\rm eff}$ & \nonvrslope $\pm$ \nonvrslopeerr & \nonvrconst $\pm$ \nonvrconsterr & \nonvrscatter $\pm$ \nonvrscattererr & \nonvrpearsonrank & N/A \\
all & $V_{W20,i}$ & $r_{\rm eff}$ & \allvrslope $\pm$ \allvrslopeerr & \allvrconst $\pm$ \allvrconsterr & \allvrscatter $\pm$ \allvrscattererr & \allvrpearsonrank & N/A \\
isolated only & $M_{\rm baryon}$ & $V_{W20,i}$ & \isomvslope $\pm$ \isomvslopeerr & \isomvconst $\pm$ \isomvconsterr & \isomvscatter $\pm$ \isomvscattererr & \isomvpearsonrank & \ref{fig_v_r_m} \\
non-isolated only & $M_{\rm baryon}$ & $V_{W20,i}$ & \nonmvslope $\pm$ \nonmvslopeerr & \nonmvconst $\pm$ \nonmvconsterr & \nonmvscatter $\pm$ \nonmvscattererr & \nonmvpearsonrank & N/A \\
all & $M_{\rm baryon}$ & $V_{W20,i}$ & \allmvslope $\pm$ \allmvslopeerr & \allmvconst $\pm$ \allmvconsterr & \allmvscatter $\pm$ \allmvscattererr & \allmvpearsonrank & N/A \\
isolated only & $V_{W20,i}$ & $M_{\rm baryon}$ & \isovmslope $\pm$ \isovmslopeerr & \isovmconst $\pm$ \isovmconsterr & \isovmscatter $\pm$ \isovmscattererr & \isovmpearsonrank & N/A \\
non-isolated only & $V_{W20,i}$ & $M_{\rm baryon}$ & \nonvmslope $\pm$ \nonvmslopeerr & \nonvmconst $\pm$ \nonvmconsterr & \nonvmscatter $\pm$ \nonvmscattererr & \nonvmpearsonrank & N/A \\
all & $V_{W20,i}$ & $M_{\rm baryon}$ & \allvmslope $\pm$ \allvmslopeerr & \allvmconst $\pm$ \allvmconsterr & \allvmscatter $\pm$ \allvmscattererr & \allvmpearsonrank & N/A
\enddata
\tablecomments{All linear relations fit in this paper. Sample, domain, range, slope, constant, scatter, Pearson r and Figure reference for each fit. For all mass-velocity-radius scaling relations, we have excluded galaxies with bars and with axis ratios greater than 0.65 from the galaxy scaling relation fits.}
\label{tbl_scaling_fits}
\end{deluxetable}

\begin{thebibliography}{133}
\expandafter\ifx\csname natexlab\endcsname\relax\def\natexlab#1{#1}\fi

\bibitem[{Abraham \& van Dokkum(2014)}]{Abraham:2014dg}
Abraham, R.~G., \& van Dokkum, P.~G. 2014, Publications of the Astronomical
  Society of the Pacific, 126, 55

\bibitem[{Aihara {et~al.}(2011)Aihara, Allende~Prieto, An, Anderson, Aubourg,
  Balbinot, Beers, Berlind, Bickerton, Bizyaev, Blanton, Bochanski, Bolton,
  Bovy, Brandt, Brinkmann, Brown, Brownstein, Busca, Campbell, Carr, Chen,
  Chiappini, Comparat, Connolly, Cortes, Croft, Cuesta, da~Costa, Davenport,
  Dawson, Dhital, Ealet, Ebelke, Edmondson, Eisenstein, Escoffier, Esposito,
  Evans, Fan, Femen{\'\i}a~Castell{\'a}, Font-Ribera, Frinchaboy, Ge,
  Gillespie, Gilmore, Gonz{\'a}lez~Hern{\'a}ndez, Gott, Gould, Grebel, Gunn,
  Hamilton, Harding, Harris, Hawley, Hearty, Ho, Hogg, Holtzman, Honscheid,
  Inada, Ivans, Jiang, Johnson, Jordan, Jordan, Kazin, Kirkby, Klaene, Knapp,
  Kneib, Kochanek, Koesterke, Kollmeier, Kron, Lampeitl, Lang, Le~Goff, Lee,
  Lin, Long, Loomis, Lucatello, Lundgren, Lupton, Ma, MacDonald, Mahadevan,
  Maia, Makler, Malanushenko, Malanushenko, Mandelbaum, Maraston, Margala,
  Masters, McBride, McGehee, McGreer, M{\'e}nard, Miralda-Escud{\'e}, Morrison,
  Mullally, Muna, Munn, Murayama, Myers, Naugle, Fausti~Neto, Nguyen, Nichol,
  O'Connell, Ogando, Olmstead, Oravetz, Padmanabhan, Palanque-Delabrouille,
  Pan, Pandey, P{\^a}ris, Percival, Petitjean, Pfaffenberger, Pforr, Phleps,
  Pichon, Pieri, Prada, Price-Whelan, Raddick, Ramos, Reyl{\'e}, Rich,
  Richards, Rix, Robin, Rocha-Pinto, Rockosi, Roe, Rollinde, Ross, Ross,
  Rossetto, S{\'a}nchez, Sayres, Schlegel, Schlesinger, Schmidt, Schneider,
  Sheldon, Shu, Simmerer, Simmons, Sivarani, Snedden, Sobeck, Steinmetz,
  Strauss, Szalay, Tanaka, Thakar, Thomas, Tinker, Tofflemire, Tojeiro,
  Tremonti, Vandenberg, Vargas~Maga{\~n}a, Verde, Vogt, Wake, Wang, Weaver,
  Weinberg, White, White, Yanny, Yasuda, Yeche, \& Zehavi}]{Aihara:2011kj}
Aihara, H., {et~al.} 2011, The Astrophysical Journal Supplement Series, 193, 29

\bibitem[{Argudo-Fern{\'a}ndez {et~al.}(2014)Argudo-Fern{\'a}ndez, Verley,
  Bergond, Sulentic, Sabater, Fern{\'a}ndez~Lorenzo, Espada, Leon,
  S{\'a}nchez-Exp{\'o}sito, Santander-Vela, \&
  Verdes-Montenegro}]{ArgudoFernandez:2014cq}
Argudo-Fern{\'a}ndez, M., {et~al.} 2014, Astronomy {\&} Astrophysics, 564, A94

\bibitem[{Avila-Reese {et~al.}(2008)Avila-Reese, Zavala, Firmani, \&
  Hern{\'a}ndez-Toledo}]{AvilaReese:2008gz}
Avila-Reese, V., Zavala, J., Firmani, C., \& Hern{\'a}ndez-Toledo, H.~M. 2008,
  The Astronomical Journal, 136, 1340

\bibitem[{Barton {et~al.}(2007)Barton, Arnold, Zentner, Bullock, \&
  Wechsler}]{Barton:2007be}
Barton, E.~J., Arnold, J.~A., Zentner, A.~R., Bullock, J.~S., \& Wechsler,
  R.~H. 2007, The Astrophysical Journal, 671, 1538

\bibitem[{Bedregal {et~al.}(2006)Bedregal, Aragon-Salamanca, \&
  Merrifield}]{Bedregal:2006jp}
Bedregal, A.~G., Aragon-Salamanca, A., \& Merrifield, M.~R. 2006, Monthly
  Notices of the Royal Astronomical Society, 373, 1125

\bibitem[{Begum {et~al.}(2008)Begum, Chengalur, Karachentsev, Sharina, \&
  Kaisin}]{Begum:2008cx}
Begum, A., Chengalur, J.~N., Karachentsev, I.~D., Sharina, M.~E., \& Kaisin,
  S.~S. 2008, Monthly Notices of the Royal Astronomical Society, 386, 1667

\bibitem[{Behroozi {et~al.}(2010)Behroozi, Conroy, \&
  Wechsler}]{Behroozi:2010ja}
Behroozi, P.~S., Conroy, C., \& Wechsler, R.~H. 2010, The Astrophysical
  Journal, 717, 379

\bibitem[{Blanton {et~al.}(2005{\natexlab{a}})Blanton, Eisenstein, Hogg,
  Schlegel, \& Brinkmann}]{Blanton:2005eb}
Blanton, M.~R., Eisenstein, D., Hogg, D.~W., Schlegel, D.~J., \& Brinkmann, J.
  2005{\natexlab{a}}, The Astrophysical Journal, 629, 143

\bibitem[{Blanton {et~al.}(2008)Blanton, Geha, \& West}]{Blanton:2008il}
Blanton, M.~R., Geha, M., \& West, A.~A. 2008, Publications of the Astronomical
  Society of the Pacific, 682, 861

\bibitem[{Blanton {et~al.}(2011)Blanton, Kazin, Muna, Weaver, \&
  Price-Whelan}]{Blanton:2011dv}
Blanton, M.~R., Kazin, E., Muna, D., Weaver, B.~A., \& Price-Whelan, A. 2011,
  The Astronomical Journal, 142, 31

\bibitem[{Blanton {et~al.}(2005{\natexlab{b}})Blanton, Lupton, Schlegel,
  Strauss, Brinkmann, Fukugita, \& Loveday}]{Blanton:2005gz}
Blanton, M.~R., Lupton, R.~H., Schlegel, D.~J., Strauss, M.~A., Brinkmann, J.,
  Fukugita, M., \& Loveday, J. 2005{\natexlab{b}}, The Astrophysical Journal,
  631, 208

\bibitem[{Blanton \& Roweis(2007)}]{Blanton:2007kl}
Blanton, M.~R., \& Roweis, S. 2007, The Astronomical Journal, 133, 734

\bibitem[{Boselli \& Gavazzi(2006)}]{Boselli:2006ei}
Boselli, A., \& Gavazzi, G. 2006, Publications of the Astronomical Society of
  the Pacific, 118, 517

\bibitem[{Broeils \& Rhee(1997)}]{Broeils:1997ut}
Broeils, A.~H., \& Rhee, M.~H. 1997, Astronomy {\&} Astrophysics, 324, 877

\bibitem[{Brooks \& Zolotov(2014)}]{Brooks:2014jv}
Brooks, A.~M., \& Zolotov, A. 2014, The Astrophysical Journal, 786, 87

\bibitem[{Brough {et~al.}(2013)Brough, Croom, Sharp, Hopkins, Taylor, Baldry,
  Gunawardhana, Liske, Norberg, Robotham, Bauer, Bland-Hawthorn, Colless,
  Foster, Kelvin, Lara-Lopez, Lopez-Sanchez, Loveday, Owers, Pimbblet, \&
  Prescott}]{Brough:2013df}
Brough, S., {et~al.} 2013, Monthly Notices of the Royal Astronomical Society,
  435, 2903

\bibitem[{Bullock {et~al.}(2001)Bullock, Kolatt, Sigad, Somerville, Kravtsov,
  Klypin, Primack, \& Dekel}]{Bullock:2001hp}
Bullock, J.~S., Kolatt, T.~S., Sigad, Y., Somerville, R.~S., Kravtsov, A.~V.,
  Klypin, A.~A., Primack, J.~R., \& Dekel, A. 2001, Monthly Notices of the
  Royal Astronomical Society, 321, 559

\bibitem[{Cannon {et~al.}(2015)Cannon, Martinkus, Leisman, Haynes, Adams,
  Giovanelli, Hallenbeck, Janowiecki, Jones, J{\'o}zsa, Koopmann, Nichols,
  Papastergis, Rhode, Salzer, \& Troischt}]{Cannon:2015cu}
Cannon, J.~M., {et~al.} 2015, The Astronomical Journal, 149, 72

\bibitem[{Casasola {et~al.}(2004)Casasola, Bettoni, \&
  Galletta}]{Casasola:2004iq}
Casasola, V., Bettoni, D., \& Galletta, G. 2004, Astronomy {\&} Astrophysics,
  422, 941

\bibitem[{Catinella {et~al.}(2010)Catinella, Schiminovich, Kauffmann, Fabello,
  Wang, Hummels, Lemonias, Moran, Wu, Giovanelli, Haynes, Heckman, Basu-Zych,
  Blanton, Brinchmann, Budav{\'a}ri, Gon{\c c}alves, Johnson, Kennicutt,
  Madore, Martin, Rich, Tacconi, Thilker, Wild, \& Wyder}]{Catinella:2010eo}
Catinella, B., {et~al.} 2010, Monthly Notices of the Royal Astronomical
  Society, 403, 683

\bibitem[{Catinella {et~al.}(2012)Catinella, Kauffmann, Schiminovich, Lemonias,
  Scannapieco, Wang, Fabello, Hummels, Moran, Wu, Cooper, Giovanelli, Haynes,
  Heckman, \& Saintonge}]{Catinella:2012ea}
---. 2012, Monthly Notices of the Royal Astronomical Society, 420, 1959

\bibitem[{Catinella {et~al.}(2013)Catinella, Schiminovich, Cortese, Fabello,
  Hummels, Moran, Lemonias, Cooper, Wu, Heckman, \& Wang}]{Catinella:2013ej}
---. 2013, Monthly Notices of the Royal Astronomical Society, 436, 34

\bibitem[{Chabrier(2003)}]{Chabrier:2003ki}
Chabrier, G. 2003, Publications of the Astronomical Society of the Pacific,
  115, 763

\bibitem[{Conroy {et~al.}(2007)Conroy, Prada, Newman, Croton, Coil, Conselice,
  Cooper, Davis, Faber, Gerke, Guhathakurta, Klypin, Koo, \&
  Yan}]{Conroy:2007ik}
Conroy, C., {et~al.} 2007, The Astrophysical Journal, 654, 153

\bibitem[{Cort{\'e}s {et~al.}(2008)Cort{\'e}s, Kenney, \&
  Hardy}]{Cortes:2008cy}
Cort{\'e}s, J.~R., Kenney, J. D.~P., \& Hardy, E. 2008, The Astrophysical
  Journal, 683, 78

\bibitem[{Cortese {et~al.}(2011)Cortese, Catinella, Boissier, Boselli, \&
  Heinis}]{Cortese:2011hz}
Cortese, L., Catinella, B., Boissier, S., Boselli, A., \& Heinis, S. 2011,
  Monthly Notices of the Royal Astronomical Society, 415, 1797

\bibitem[{Cortese {et~al.}(2012)Cortese, Ciesla, Boselli, Bianchi, Gomez,
  Smith, Bendo, Eales, Pohlen, Baes, Corbelli, Davies, Hughes, Hunt, Madden,
  Pierini, di~Serego~Alighieri, Zibetti, Boquien, Clements, Cooray, Galametz,
  Magrini, Pappalardo, Spinoglio, \& Vlahakis}]{Cortese:2012cv}
Cortese, L., {et~al.} 2012, Astronomy {\&} Astrophysics, 540, A52

\bibitem[{Courteau {et~al.}(2007)Courteau, Dutton, van~den Bosch, MacArthur,
  Dekel, McIntosh, \& Dale}]{Courteau:2007ek}
Courteau, S., Dutton, A.~A., van~den Bosch, F.~C., MacArthur, L.~A., Dekel, A.,
  McIntosh, D.~H., \& Dale, D.~A. 2007, The Astrophysical Journal, 671, 203

\bibitem[{Croton {et~al.}(2005)Croton, Farrar, Norberg, Colless, Peacock,
  Baldry, Baugh, Bland-Hawthorn, Bridges, Cannon, Cole, Collins, Couch, Dalton,
  De~Propris, Driver, Efstathiou, Ellis, Frenk, Glazebrook, Jackson, Lahav,
  Lewis, Lumsden, Maddox, Madgwick, Peterson, Sutherland, Taylor, \& {The
  2dFGRS Team}}]{Croton:2005cl}
Croton, D.~J., {et~al.} 2005, Monthly Notices of the Royal Astronomical
  Society, 356, 1155

\bibitem[{Dalcanton {et~al.}(1997)Dalcanton, Spergel, Gunn, Schmidt, \&
  Schneider}]{Dalcanton:1997kf}
Dalcanton, J.~J., Spergel, D.~N., Gunn, J.~E., Schmidt, M., \& Schneider, D.~P.
  1997, The Astronomical Journal, 114, 635

\bibitem[{Davis {et~al.}(2011)Davis, Bureau, Young, Alatalo, Blitz, Cappellari,
  Scott, Bois, Bournaud, Davies, de~Zeeuw, Emsellem, Khochfar, Krajnovi{\'c},
  Kuntschner, Lablanche, McDermid, Morganti, Naab, Oosterloo, Sarzi, Serra, \&
  Weijmans}]{Davis:2011bg}
Davis, T.~A., {et~al.} 2011, Monthly Notices of the Royal Astronomical Society,
  414, 968

\bibitem[{de~Blok \& Walter(2014)}]{deBlok:2014bm}
de~Blok, W. J.~G., \& Walter, F. 2014, The Astronomical Journal, 147, 96

\bibitem[{Draine(2011)}]{Draine:2011tr}
Draine, B.~T. 2011, Physics of the Interstellar and Intergalactic Medium by
  Bruce T. Draine. Princeton University Press, 2011. ISBN: 978-0-691-12214-4,
  -1

\bibitem[{Dressler(1980)}]{Dressler:1980ie}
Dressler, A. 1980, The Astrophysical Journal, 236, 351

\bibitem[{Dutton {et~al.}(2007)Dutton, van~den Bosch, Dekel, \&
  Courteau}]{Dutton:2007he}
Dutton, A.~A., van~den Bosch, F.~C., Dekel, A., \& Courteau, S. 2007, The
  Astrophysical Journal, 654, 27

\bibitem[{Dutton {et~al.}(2010)Dutton, Bosch, Faber, Simard, Kassin, Koo,
  Bundy, Huang, Weiner, Cooper, Newman, Mozena, \& Koekemoer}]{Dutton:2010ci}
Dutton, A.~A., {et~al.} 2010, Monthly Notices of the Royal Astronomical
  Society, 410, no

\bibitem[{Fern{\'a}ndez~Lorenzo {et~al.}(2013)Fern{\'a}ndez~Lorenzo, Sulentic,
  Verdes-Montenegro, \& Argudo-Fern{\'a}ndez}]{FernandezLorenzo:2013ez}
Fern{\'a}ndez~Lorenzo, M., Sulentic, J., Verdes-Montenegro, L., \&
  Argudo-Fern{\'a}ndez, M. 2013, Monthly Notices of the Royal Astronomical
  Society, 434, 325

\bibitem[{Ferrero {et~al.}(2012)Ferrero, Abadi, Navarro, Sales, \&
  Gurovich}]{Ferrero:2012bt}
Ferrero, I., Abadi, M.~G., Navarro, J.~F., Sales, L.~V., \& Gurovich, S. 2012,
  Monthly Notices of the Royal Astronomical Society, 425, 2817

\bibitem[{Firmani \& Avila-Reese(2009)}]{Firmani:2009ho}
Firmani, C., \& Avila-Reese, V. 2009, Monthly Notices of the Royal Astronomical
  Society, 396, 1675

\bibitem[{Gallazzi {et~al.}(2008)Gallazzi, Bell, Wolf, Gray, Papovich, Barden,
  Peng, Meisenheimer, Heymans, van Kampen, Gilmour, Balogh, McIntosh, Bacon,
  Barazza, B{\"o}hm, Caldwell, H{\"a}u{\ss}ler, Jahnke, Jogee, Lane, Robaina,
  Sanchez, Taylor, Wisotzki, \& Zheng}]{Gallazzi:2008ku}
Gallazzi, A., {et~al.} 2008, The Astrophysical Journal, 690, 1883

\bibitem[{Geha {et~al.}(2006)Geha, Blanton, Masjedi, \& West}]{Geha:2006jx}
Geha, M., Blanton, M.~R., Masjedi, M., \& West, A.~A. 2006, Publications of the
  Astronomical Society of the Pacific, 653, 240

\bibitem[{Geha {et~al.}(2012)Geha, Blanton, Yan, \& Tinker}]{Geha:2012eu}
Geha, M., Blanton, M.~R., Yan, R., \& Tinker, J.~L. 2012, The Astrophysical
  Journal, 757, 85

\bibitem[{Gnedin(2012)}]{Gnedin:2012ip}
Gnedin, N.~Y. 2012, The Astrophysical Journal, 754, 113

\bibitem[{Grcevich \& Putman(2009)}]{Grcevich:2009fs}
Grcevich, J., \& Putman, M.~E. 2009, The Astrophysical Journal, 696, 385

\bibitem[{Gunn \& Gott(1972)}]{Gunn:1972gx}
Gunn, J.~E., \& Gott, J. R.~I. 1972, The Astrophysical Journal, 176, 1

\bibitem[{Hall {et~al.}(2012)Hall, Courteau, Dutton, McDonald, \&
  Zhu}]{Hall:2012hh}
Hall, M., Courteau, S., Dutton, A.~A., McDonald, M., \& Zhu, Y. 2012, Monthly
  Notices of the Royal Astronomical Society, 425, 2741

\bibitem[{Hallenbeck {et~al.}(2014)Hallenbeck, Huang, Spekkens, Haynes,
  Giovanelli, Adams, Brinchmann, Chengalur, Hunt, Masters, \&
  Saintonge}]{Hallenbeck:2014ku}
Hallenbeck, G., {et~al.} 2014, The Astronomical Journal, 148, 69

\bibitem[{Haynes \& Giovanelli(1984)}]{Haynes:1984el}
Haynes, M.~P., \& Giovanelli, R. 1984, The Astronomical Journal, 89, 758

\bibitem[{Haynes {et~al.}(2011)Haynes, Giovanelli, Martin, Hess, Saintonge,
  Adams, Hallenbeck, Hoffman, Huang, Kent, Koopmann, Papastergis, Stierwalt,
  Balonek, Craig, Higdon, Kornreich, Miller, O'Donoghue, Olowin, Rosenberg,
  Spekkens, Troischt, \& Wilcots}]{Haynes:2011en}
Haynes, M.~P., {et~al.} 2011, The Astronomical Journal, 142, 170

\bibitem[{Hinz {et~al.}(2003)Hinz, Rieke, \& Caldwell}]{Hinz:2003bb}
Hinz, J.~L., Rieke, G.~H., \& Caldwell, N. 2003, The Astronomical Journal, 126,
  2622

\bibitem[{Hoyle {et~al.}(2011)Hoyle, Masters, Nichol, Edmondson, Smith,
  Lintott, Scranton, Bamford, Schawinski, \& Thomas}]{Hoyle:2011bk}
Hoyle, B., {et~al.} 2011, Monthly Notices of the Royal Astronomical Society,
  415, 3627

\bibitem[{Huang {et~al.}(2012{\natexlab{a}})Huang, Haynes, Giovanelli, \&
  Brinchmann}]{Huang:2012gk}
Huang, S., Haynes, M.~P., Giovanelli, R., \& Brinchmann, J. 2012{\natexlab{a}},
  The Astrophysical Journal, 756, 113

\bibitem[{Huang {et~al.}(2012{\natexlab{b}})Huang, Haynes, Giovanelli,
  Brinchmann, Stierwalt, \& Neff}]{Huang:2012bv}
Huang, S., Haynes, M.~P., Giovanelli, R., Brinchmann, J., Stierwalt, S., \&
  Neff, S.~G. 2012{\natexlab{b}}, The Astronomical Journal, 143, 133

\bibitem[{Hunter(2007)}]{Hunter:2007ih}
Hunter, J.~D. 2007, Computing in Science {\&} Engineering, 9, 90

\bibitem[{Jarrett {et~al.}(2000)Jarrett, Chester, Cutri, Schneider, Skrutskie,
  \& Huchra}]{Jarrett:2000fz}
Jarrett, T.~H., Chester, T., Cutri, R., Schneider, S., Skrutskie, M., \&
  Huchra, J.~P. 2000, The Astronomical Journal, 119, 2498

\bibitem[{Karachentsev {et~al.}(2011)Karachentsev, Makarov, Karachentseva, \&
  Melnyk}]{Karachentsev:2011ju}
Karachentsev, I.~D., Makarov, D.~I., Karachentseva, V.~E., \& Melnyk, O.~V.
  2011, Astrophysical Bulletin, 66, 1

\bibitem[{Karachentseva(1973)}]{Karachentseva:1973we}
Karachentseva, V.~E. 1973, Soobshcheniya Spetsial'noj Astrofizicheskoj
  Observatorii, 8, 3

\bibitem[{Kassin {et~al.}(2014)Kassin, Brooks, Governato, Weiner, \&
  Gardner}]{Kassin:2014jp}
Kassin, S.~A., Brooks, A., Governato, F., Weiner, B.~J., \& Gardner, J.~P.
  2014, The Astrophysical Journal, 790, 89

\bibitem[{Kassin {et~al.}(2007)Kassin, Weiner, Faber, Koo, Lotz, Diemand,
  Harker, Bundy, Metevier, Phillips, Cooper, Croton, Konidaris, Noeske, \&
  Willmer}]{Kassin:2007ey}
Kassin, S.~A., {et~al.} 2007, arXiv.org, L35

\bibitem[{Kauffmann(2014)}]{Kauffmann:2014vf}
Kauffmann, G. 2014, eprint arXiv:1401.8091

\bibitem[{Kauffmann {et~al.}(2004)Kauffmann, White, Heckman, M{\'e}nard,
  Brinchmann, Charlot, Tremonti, \& Brinkmann}]{Kauffmann:2004cw}
Kauffmann, G., White, S. D.~M., Heckman, T.~M., M{\'e}nard, B., Brinchmann, J.,
  Charlot, S., Tremonti, C., \& Brinkmann, J. 2004, Monthly Notices of the
  Royal Astronomical Society, 353, 713

\bibitem[{Kelly(2007)}]{Kelly:2007bv}
Kelly, B.~C. 2007, The Astrophysical Journal, 665, 1489

\bibitem[{Kenney {et~al.}(2013)Kenney, Geha, J{\'a}chym, Crowl, Dague, Chung,
  van Gorkom, \& Vollmer}]{Kenney:2013bf}
Kenney, J. D.~P., Geha, M., J{\'a}chym, P., Crowl, H.~H., Dague, W., Chung, A.,
  van Gorkom, J., \& Vollmer, B. 2013, The Astrophysical Journal, 780, 119

\bibitem[{Kenney \& Young(1989)}]{Kenney:1989dg}
Kenney, J. D.~P., \& Young, J.~S. 1989, The Astrophysical Journal, 344, 171

\bibitem[{Kirby {et~al.}(2014)Kirby, Bullock, Boylan-Kolchin, Kaplinghat, \&
  Cohen}]{Kirby:2014gk}
Kirby, E.~N., Bullock, J.~S., Boylan-Kolchin, M., Kaplinghat, M., \& Cohen,
  J.~G. 2014, Monthly Notices of the Royal Astronomical Society, 439, 1015

\bibitem[{Klypin {et~al.}(2014)Klypin, Karachentsev, Makarov, \&
  Nasonova}]{Klypin:2014ue}
Klypin, A., Karachentsev, I., Makarov, D., \& Nasonova, O. 2014, eprint
  arXiv:1405.4523

\bibitem[{Klypin {et~al.}(2011)Klypin, Trujillo-Gomez, \&
  Primack}]{Klypin:2011bd}
Klypin, A.~A., Trujillo-Gomez, S., \& Primack, J. 2011, The Astrophysical
  Journal, 740, 102

\bibitem[{Komatsu {et~al.}(2011)Komatsu, Smith, Dunkley, Bennett, Gold,
  Hinshaw, Jarosik, Larson, Nolta, Page, Spergel, Halpern, Hill, Kogut, Limon,
  Meyer, Odegard, Tucker, Weiland, Wollack, \& Wright}]{Komatsu:2011in}
Komatsu, E., {et~al.} 2011, The Astrophysical Journal Supplement Series, 192,
  18

\bibitem[{Koribalski {et~al.}(2004)Koribalski, Staveley-Smith, Kilborn, Ryder,
  Kraan-Korteweg, Ryan-Weber, Ekers, Jerjen, Henning, Putman, Zwaan, de~Blok,
  Calabretta, Disney, Minchin, Bhathal, Boyce, Drinkwater, Freeman, Gibson,
  Green, Haynes, Juraszek, Kesteven, Knezek, Mader, Marquarding, Meyer, Mould,
  Oosterloo, O'Brien, Price, Sadler, Schr{\"o}der, Stewart, Stootman, Waugh,
  Warren, Webster, \& Wright}]{Koribalski:2004cv}
Koribalski, B.~S., {et~al.} 2004, The Astronomical Journal, 128, 16

\bibitem[{Kravtsov {et~al.}(2004)Kravtsov, Gnedin, \& Klypin}]{Kravtsov:2004he}
Kravtsov, A.~V., Gnedin, O.~Y., \& Klypin, A.~A. 2004, The Astrophysical
  Journal, 609, 482

\bibitem[{Kreckel {et~al.}(2012)Kreckel, Platen, Arag{\'o}n-Calvo, van Gorkom,
  van~de Weygaert, van~der Hulst, \& Beygu}]{Kreckel:2012it}
Kreckel, K., Platen, E., Arag{\'o}n-Calvo, M.~A., van Gorkom, J.~H., van~de
  Weygaert, R., van~der Hulst, J.~M., \& Beygu, B. 2012, The Astronomical
  Journal, 144, 16

\bibitem[{Leroy {et~al.}(2008)Leroy, Walter, Brinks, Bigiel, de~Blok, Madore,
  \& Thornley}]{Leroy:2008jk}
Leroy, A.~K., Walter, F., Brinks, E., Bigiel, F., de~Blok, W. J.~G., Madore,
  B., \& Thornley, M.~D. 2008, The Astronomical Journal, 136, 2782

\bibitem[{Leroy {et~al.}(2009)Leroy, Walter, Bigiel, Usero, Weiss, Brinks,
  de~Blok, Kennicutt, Schuster, Kramer, Wiesemeyer, \& Roussel}]{Leroy:2009di}
Leroy, A.~K., {et~al.} 2009, The Astronomical Journal, 137, 4670

\bibitem[{Lisenfeld {et~al.}(2011)Lisenfeld, Espada, Verdes-Montenegro, Kuno,
  Leon, Sabater, Sato, Sulentic, Verley, \& Yun}]{Lisenfeld:2011es}
Lisenfeld, U., {et~al.} 2011, Astronomy {\&} Astrophysics, 534, A102

\bibitem[{Maddox {et~al.}(2014)Maddox, Hess, Obreschkow, Jarvis, \&
  Blyth}]{Maddox:2014fp}
Maddox, N., Hess, K.~M., Obreschkow, D., Jarvis, M.~J., \& Blyth, S.~L. 2014,
  Monthly Notices of the Royal Astronomical Society, 447, 1610

\bibitem[{Martin {et~al.}(2005)Martin, Fanson, Schiminovich, Morrissey,
  Friedman, Barlow, Conrow, Grange, Jelinsky, Milliard, Siegmund, Bianchi,
  Byun, Donas, Forster, Heckman, Lee, Madore, Malina, Neff, Rich, Small,
  Surber, Szalay, Welsh, \& Wyder}]{Martin:2005ko}
Martin, D.~C., {et~al.} 2005, The Astrophysical Journal, 619, L1

\bibitem[{McGaugh {et~al.}(2000)McGaugh, Schombert, Bothun, \&
  de~Blok}]{McGaugh:2000hx}
McGaugh, S., Schombert, J., Bothun, G., \& de~Blok, E. 2000, arXiv.org, L99

\bibitem[{McGaugh(2005)}]{McGaugh:2005bc}
McGaugh, S.~S. 2005, The Astrophysical Journal, 632, 859

\bibitem[{McGaugh(2012)}]{McGaugh:2012ev}
---. 2012, The Astronomical Journal, 143, 40

\bibitem[{McGaugh \& de~Blok(1997)}]{McGaugh:1997fk}
McGaugh, S.~S., \& de~Blok, W. J.~G. 1997, The Astrophysical Journal, 481, 689

\bibitem[{McGaugh \& Schombert(2015)}]{McGaugh:2015eu}
McGaugh, S.~S., \& Schombert, J.~M. 2015, The Astrophysical Journal, 802, 18

\bibitem[{McGaugh {et~al.}(2009)McGaugh, Schombert, de~Blok, \&
  Zagursky}]{McGaugh:2009by}
McGaugh, S.~S., Schombert, J.~M., de~Blok, W. J.~G., \& Zagursky, M.~J. 2009,
  The Astrophysical Journal, 708, L14

\bibitem[{McGaugh \& Wolf(2010)}]{McGaugh:2010ky}
McGaugh, S.~S., \& Wolf, J. 2010, The Astrophysical Journal, 722, 248

\bibitem[{Merritt {et~al.}(2014)Merritt, van Dokkum, \&
  Abraham}]{Merritt:2014ha}
Merritt, A., van Dokkum, P., \& Abraham, R. 2014, The Astrophysical Journal,
  787, L37

\bibitem[{Mo {et~al.}(2010)Mo, van~den Bosch, \& White}]{Mo:2010wea}
Mo, H., van~den Bosch, F.~C., \& White, S. 2010, Galaxy Formation and
  Evolution, by Houjun Mo , Frank van den Bosch , Simon White, Cambridge, UK:
  Cambridge University Press, 2010, -1

\bibitem[{Moore {et~al.}(1996)Moore, Katz, Lake, Dressler, \&
  Oemler}]{Moore:1996il}
Moore, B., Katz, N., Lake, G., Dressler, A., \& Oemler, A. 1996, Nature, 379,
  613

\bibitem[{Muldrew {et~al.}(2011)Muldrew, Croton, Skibba, Pearce, Ann, Baldry,
  Brough, Choi, Conselice, Cowan, Gallazzi, Gray, Gr{\"u}tzbauch, Li, Park,
  Pilipenko, Podgorzec, Robotham, Wilman, Yang, Zhang, \&
  Zibetti}]{Muldrew:2011go}
Muldrew, S.~I., {et~al.} 2011, Monthly Notices of the Royal Astronomical
  Society, 419, 2670

\bibitem[{Neistein {et~al.}(1999)Neistein, Maoz, Rix, \&
  Tonry}]{Neistein:1999io}
Neistein, E., Maoz, D., Rix, H.-W., \& Tonry, J.~L. 1999, The Astronomical
  Journal, 117, 2666

\bibitem[{Papastergis {et~al.}(2012)Papastergis, Cattaneo, Huang, Giovanelli,
  \& Haynes}]{Papastergis:2012cb}
Papastergis, E., Cattaneo, A., Huang, S., Giovanelli, R., \& Haynes, M.~P.
  2012, The Astrophysical Journal, 759, 138

\bibitem[{Papastergis {et~al.}(2015)Papastergis, Giovanelli, Haynes, \&
  Shankar}]{Papastergis:2015bc}
Papastergis, E., Giovanelli, R., Haynes, M.~P., \& Shankar, F. 2015, Astronomy
  {\&} Astrophysics, 574, A113

\bibitem[{Pe{\~n}arrubia {et~al.}(2008)Pe{\~n}arrubia, Navarro, \&
  McConnachie}]{Penarrubia:2008eq}
Pe{\~n}arrubia, J., Navarro, J.~F., \& McConnachie, A.~W. 2008, The
  Astrophysical Journal, 673, 226

\bibitem[{Peng {et~al.}(2010)Peng, Lilly, Kova{\v c}, Bolzonella, Pozzetti,
  Renzini, Zamorani, Ilbert, Knobel, Iovino, Maier, Cucciati, Tasca, Carollo,
  Silverman, Kampczyk, de~Ravel, Sanders, Scoville, Contini, Mainieri,
  Scodeggio, Kneib, Le~F{\`e}vre, Bardelli, Bongiorno, Caputi, Coppa, de~la
  Torre, Franzetti, Garilli, Lamareille, Le~Borgne, Le~Brun, Mignoli, Montero,
  Pello, Ricciardelli, Tanaka, Tresse, Vergani, Welikala, Zucca, Oesch, Abbas,
  Barnes, Bordoloi, Bottini, Cappi, Cassata, Cimatti, Fumana, Hasinger,
  Koekemoer, Leauthaud, Maccagni, Marinoni, McCracken, Memeo, Meneux, Nair,
  Porciani, Presotto, \& Scaramella}]{Peng:2010gn}
Peng, Y.-j., {et~al.} 2010, The Astrophysical Journal, 721, 193

\bibitem[{Perea {et~al.}(1997)Perea, del Olmo, Verdes-Montenegro, \&
  Yun}]{Perea:1997vp}
Perea, J., del Olmo, A., Verdes-Montenegro, L., \& Yun, M.~S. 1997, ApJ, 490,
  166

\bibitem[{Reines {et~al.}(2013)Reines, Greene, \& Geha}]{Reines:2013bp}
Reines, A.~E., Greene, J.~E., \& Geha, M. 2013, The Astrophysical Journal, 775,
  116

\bibitem[{Roychowdhury {et~al.}(2014)Roychowdhury, Chengalur, Kaisin, \&
  Karachentsev}]{Roychowdhury:2014jf}
Roychowdhury, S., Chengalur, J.~N., Kaisin, S.~S., \& Karachentsev, I.~D. 2014,
  Monthly Notices of the Royal Astronomical Society, 445, 1392

\bibitem[{Saintonge {et~al.}(2011)Saintonge, Kauffmann, Kramer, Tacconi,
  Buchbender, Catinella, Fabello, Graci{\'a}-Carpio, Wang, Cortese, Fu, Genzel,
  Giovanelli, Guo, Haynes, Heckman, Krumholz, Lemonias, Li, Moran,
  Rodriguez-Fernandez, Schiminovich, Schuster, \& Sievers}]{Saintonge:2011hz}
Saintonge, A., {et~al.} 2011, Monthly Notices of the Royal Astronomical
  Society, 415, 32

\bibitem[{Schombert {et~al.}(2001)Schombert, McGaugh, \&
  Eder}]{Schombert:2001gc}
Schombert, J.~M., McGaugh, S.~S., \& Eder, J.~A. 2001, The Astronomical
  Journal, 121, 2420

\bibitem[{Schruba {et~al.}(2012)Schruba, Leroy, Walter, Bigiel, Brinks,
  de~Blok, Kramer, Rosolowsky, Sandstrom, Schuster, Usero, Weiss, \&
  Wiesemeyer}]{Schruba:2012bb}
Schruba, A., {et~al.} 2012, The Astronomical Journal, 143, 138

\bibitem[{Singhal(2008)}]{Singhal:2008tp}
Singhal, A. 2008, Proquest Dissertations And Theses 2008. Section 0246, Part
  0606 222 pages; [Ph.D. dissertation].United States -- Virginia: University of
  Virginia; 2008. Publication Number: AAT 3300266. Source: DAI-B 69/01, Jul
  2008

\bibitem[{Spekkens {et~al.}(2014)Spekkens, Urbancic, Mason, Willman, \&
  Aguirre}]{Spekkens:2014he}
Spekkens, K., Urbancic, N., Mason, B.~S., Willman, B., \& Aguirre, J.~E. 2014,
  The Astrophysical Journal, 795, L5

\bibitem[{Springob {et~al.}(2005)Springob, Haynes, Giovanelli, \&
  Kent}]{Springob:2005db}
Springob, C.~M., Haynes, M.~P., Giovanelli, R., \& Kent, B.~R. 2005, The
  Astrophysical Journal Supplement Series, 160, 149

\bibitem[{Stark {et~al.}(2009)Stark, McGaugh, \& Swaters}]{Stark:2009ks}
Stark, D.~V., McGaugh, S.~S., \& Swaters, R.~A. 2009, The Astronomical Journal,
  138, 392

\bibitem[{Stierwalt {et~al.}(2015)Stierwalt, Besla, Patton, Johnson,
  Kallivayalil, Putman, Privon, \& Ross}]{Stierwalt:2015ju}
Stierwalt, S., Besla, G., Patton, D., Johnson, K., Kallivayalil, N., Putman,
  M., Privon, G., \& Ross, G. 2015, The Astrophysical Journal, 805, 2

\bibitem[{Swaters {et~al.}(2009)Swaters, Sancisi, van Albada, \& van~der
  Hulst}]{Swaters:2009ku}
Swaters, R.~A., Sancisi, R., van Albada, T.~S., \& van~der Hulst, J.~M. 2009,
  Astronomy {\&} Astrophysics, 493, 871

\bibitem[{Tal {et~al.}(2014)Tal, Dekel, Oesch, Muzzin, Brammer, van Dokkum,
  Franx, Illingworth, Leja, Magee, Marchesini, Momcheva, Nelson, Patel, Quadri,
  Rix, Skelton, Wake, \& Whitaker}]{Tal:2014uf}
Tal, T., {et~al.} 2014, The Astrophysical Journal, 789, 164

\bibitem[{Tinker {et~al.}(2011)Tinker, Wetzel, \& Conroy}]{Tinker:2011ul}
Tinker, J., Wetzel, A., \& Conroy, C. 2011, eprint arXiv:1107.5046

\bibitem[{Tollerud {et~al.}(2011)Tollerud, Boylan-Kolchin, Barton, Bullock, \&
  Trinh}]{Tollerud:2011jn}
Tollerud, E.~J., Boylan-Kolchin, M., Barton, E.~J., Bullock, J.~S., \& Trinh,
  C.~Q. 2011, The Astrophysical Journal, 738, 102

\bibitem[{Tollerud {et~al.}(2010)Tollerud, Bullock, Graves, \&
  Wolf}]{Tollerud:2010bq}
Tollerud, E.~J., Bullock, J.~S., Graves, G.~J., \& Wolf, J. 2010, The
  Astrophysical Journal, 726, 108

\bibitem[{Toomre \& Toomre(1972)}]{Toomre:1972ji}
Toomre, A., \& Toomre, J. 1972, The Astrophysical Journal, 178, 623

\bibitem[{Toribio {et~al.}(2011)Toribio, Solanes, Giovanelli, Haynes, \&
  Masters}]{Toribio:2011eb}
Toribio, M.~C., Solanes, J.~M., Giovanelli, R., Haynes, M.~P., \& Masters,
  K.~L. 2011, The Astrophysical Journal, 732, 92

\bibitem[{Tully \& Fisher(1977)}]{Tully:1977wu}
Tully, R.~B., \& Fisher, J.~R. 1977, Astronomy {\&} Astrophysics, 54, 661

\bibitem[{van~den Bosch(2000)}]{vandenBosch:2000iw}
van~den Bosch, F.~C. 2000, The Astrophysical Journal, 530, 177

\bibitem[{van~den Bosch(2002)}]{vandenBosch:2002kk}
---. 2002, Monthly Notices of the Royal Astronomical Society, 332, 456

\bibitem[{van Dokkum {et~al.}(2015{\natexlab{a}})van Dokkum, Abraham, Merritt,
  Zhang, Geha, \& Conroy}]{vanDokkum:2015ks}
van Dokkum, P.~G., Abraham, R., Merritt, A., Zhang, J., Geha, M., \& Conroy, C.
  2015{\natexlab{a}}, The Astrophysical Journal, 798, L45

\bibitem[{van Dokkum {et~al.}(2015{\natexlab{b}})van Dokkum, Romanowsky,
  Abraham, Brodie, Conroy, Geha, Merritt, Villaume, \&
  Zhang}]{vanDokkum:2015vc}
van Dokkum, P.~G., {et~al.} 2015{\natexlab{b}}, ApJ, 804, L26

\bibitem[{van Zee(2000)}]{vanZee:2000bp}
van Zee, L. 2000, The Astronomical Journal, 119, 2757

\bibitem[{van Zee(2001)}]{vanZee:2001fm}
---. 2001, The Astronomical Journal, 121, 2003

\bibitem[{Verheijen(1997)}]{Verheijen:1997vb}
Verheijen, M. A.~W. 1997, PhD thesis, Univ. Groningen, The Netherlands , (1997)

\bibitem[{Wang {et~al.}(2014)Wang, Fu, Aumer, Kauffmann, Jozsa, Serra, Huang,
  Brinchmann, van~der Hulst, \& Bigiel}]{Wang:2014gb}
Wang, J., {et~al.} 2014, Monthly Notices of the Royal Astronomical Society,
  441, 2159

\bibitem[{Weiner {et~al.}(2006)Weiner, Willmer, Faber, Melbourne, Kassin,
  Phillips, Harker, Metevier, Vogt, \& Koo}]{Weiner:2006cz}
Weiner, B.~J., {et~al.} 2006, The Astrophysical Journal, 653, 1027

\bibitem[{Welch {et~al.}(2010)Welch, Sage, \& Young}]{Welch:2010in}
Welch, G.~A., Sage, L.~J., \& Young, L.~M. 2010, The Astrophysical Journal,
  725, 100

\bibitem[{Wetzel {et~al.}(2014)Wetzel, Tinker, Conroy, \&
  Bosch}]{Wetzel:2014ju}
Wetzel, A.~R., Tinker, J.~L., Conroy, C., \& Bosch, F. C. v.~d. 2014, Monthly
  Notices of the Royal Astronomical Society, 439, 2687

\bibitem[{Wheeler {et~al.}(2014)Wheeler, Phillips, Cooper, Boylan-Kolchin, \&
  Bullock}]{Wheeler:2014kz}
Wheeler, C., Phillips, J.~I., Cooper, M.~C., Boylan-Kolchin, M., \& Bullock,
  J.~S. 2014, Monthly Notices of the Royal Astronomical Society, 442, 1396

\bibitem[{Williams {et~al.}(2010)Williams, Bureau, \&
  Cappellari}]{Williams:2010hn}
Williams, M.~J., Bureau, M., \& Cappellari, M. 2010, Monthly Notices of the
  Royal Astronomical Society, 409, 1330

\bibitem[{Willick {et~al.}(1997)Willick, Courteau, Faber, Burstein, Dekel, \&
  Strauss}]{Willick:1997gg}
Willick, J.~A., Courteau, S., Faber, S.~M., Burstein, D., Dekel, A., \&
  Strauss, M.~A. 1997, The Astrophysical Journal Supplement Series, 109, 333

\bibitem[{Yan(2011)}]{Yan:2011in}
Yan, R. 2011, The Astronomical Journal, 142, 153

\bibitem[{Yan \& Blanton(2012)}]{Yan:2012jn}
Yan, R., \& Blanton, M.~R. 2012, The Astrophysical Journal, 747, 61

\bibitem[{Young \& Scoville(1991)}]{Young:1991hh}
Young, J.~S., \& Scoville, N.~Z. 1991, Annual Review of Astronomy and
  Astrophysics, 29, 581

\bibitem[{Yuan \& Zhu(2004)}]{Yuan:2004kr}
Yuan, Q.-r., \& Zhu, C.-x. 2004, Chinese Astronomy and Astrophysics, 28, 127

\bibitem[{Zaritsky {et~al.}(2014)Zaritsky, Courtois, Mu{\~n}oz-Mateos, Sorce,
  Erroz-Ferrer, Comer{\'o}n, Gadotti, Gil~de Paz, Hinz, Laurikainen, Kim,
  Laine, Men{\'e}ndez-Delmestre, Mizusawa, Regan, Salo, Seibert, Sheth,
  Athanassoula, Bosma, Cisternas, Ho, \& Holwerda}]{Zaritsky:2014fr}
Zaritsky, D., {et~al.} 2014, The Astronomical Journal, 147, 134

\bibitem[{Zhang {et~al.}(2009)Zhang, Li, Kauffmann, Zou, Catinella, Shen, Guo,
  \& Chang}]{Zhang:2009cr}
Zhang, W., Li, C., Kauffmann, G., Zou, H., Catinella, B., Shen, S., Guo, Q., \&
  Chang, R. 2009, Monthly Notices of the Royal Astronomical Society, 397, 1243

\bibitem[{Zolotov {et~al.}(2012)Zolotov, Brooks, Willman, Governato, Pontzen,
  Christensen, Dekel, Quinn, Shen, \& Wadsley}]{Zolotov:2012hi}
Zolotov, A., {et~al.} 2012, The Astrophysical Journal, 761, 71

\end{thebibliography}
\end{document}